\documentclass[aps,prb,superscriptaddress,twocolumn,floatfix]{revtex4-2}
\usepackage[utf8]{inputenc}
\usepackage{amsmath}
\usepackage{amsfonts}
\usepackage{amssymb}
\usepackage[dvipsnames]{xcolor}
\usepackage{graphicx}
\usepackage{physics}
\usepackage{textcomp}
\usepackage{dsfont}
\usepackage{lipsum}
\usepackage{bbold}
\usepackage{tabularx}
\usepackage{longtable}
\usepackage{array}
\usepackage{hhline}
\usepackage{upgreek}
\emergencystretch=\maxdimen
\allowdisplaybreaks

\begin{document}

\title{Control of dephasing in spin qubits during coherent transport in silicon}

\author{MengKe Feng}
\email[Corresponding Author: ]{mengke.feng@unsw.edu.au}
\affiliation{School of Electrical Engineering and Telecommunications, The University of New South Wales, Sydney, NSW 2052, Australia}
\author{Jun Yoneda}
\altaffiliation{Present Address: Tokyo Tech Academy for Super Smart Society, Tokyo Institute of Technology, Tokyo, 152-8552 Japan}
\affiliation{School of Electrical Engineering and Telecommunications, The University of New South Wales, Sydney, NSW 2052, Australia}
\author{Wister Huang}
\altaffiliation{Present Address: Otto-Stern-Weg 1, 8093 Zürich, Switzerland}
\affiliation{School of Electrical Engineering and Telecommunications, The University of New South Wales, Sydney, NSW 2052, Australia}
\author{Yue Su}
\affiliation{School of Electrical Engineering and Telecommunications, The University of New South Wales, Sydney, NSW 2052, Australia}
\author{Tuomo Tanttu}
\affiliation{School of Electrical Engineering and Telecommunications, The University of New South Wales, Sydney, NSW 2052, Australia}
\affiliation{Diraq, Sydney, New South Wales, Australia}
\author{Chih Hwan Yang}
\affiliation{School of Electrical Engineering and Telecommunications, The University of New South Wales, Sydney, NSW 2052, Australia}
\affiliation{Diraq, Sydney, New South Wales, Australia}
\author{Jesus D. Cifuentes}
\affiliation{School of Electrical Engineering and Telecommunications, The University of New South Wales, Sydney, NSW 2052, Australia}
\author{Kok Wai Chan}
\affiliation{School of Electrical Engineering and Telecommunications, The University of New South Wales, Sydney, NSW 2052, Australia}
\affiliation{Diraq, Sydney, New South Wales, Australia}
\author{William Gilbert}
\affiliation{School of Electrical Engineering and Telecommunications, The University of New South Wales, Sydney, NSW 2052, Australia}
\affiliation{Diraq, Sydney, New South Wales, Australia}
\author{Ross C. C. Leon}
\affiliation{School of Electrical Engineering and Telecommunications, The University of New South Wales, Sydney, NSW 2052, Australia}
\author{Fay E. Hudson}
\affiliation{School of Electrical Engineering and Telecommunications, The University of New South Wales, Sydney, NSW 2052, Australia}
\affiliation{Diraq, Sydney, New South Wales, Australia}
\author{Kohei M. Itoh}
\affiliation{School of Fundamental Science and Technology, Keio University, Yokohama, Japan}
\author{Arne Laucht}
\affiliation{School of Electrical Engineering and Telecommunications, The University of New South Wales, Sydney, NSW 2052, Australia}
\affiliation{Diraq, Sydney, New South Wales, Australia}
\author{Andrew S. Dzurak}
\affiliation{School of Electrical Engineering and Telecommunications, The University of New South Wales, Sydney, NSW 2052, Australia}
\affiliation{Diraq, Sydney, New South Wales, Australia}
\author{Andre Saraiva}
\email[Corresponding Author: ]{a.saraiva@unsw.edu.au}
\affiliation{School of Electrical Engineering and Telecommunications, The University of New South Wales, Sydney, NSW 2052, Australia}
\affiliation{Diraq, Sydney, New South Wales, Australia}

\begin{abstract}
    One of the key pathways towards scalability of spin-based quantum computing systems lies in achieving long-range interactions between electrons and increasing their inter-connectivity. Coherent spin transport is one of the most promising strategies to achieve this architectural advantage. Experimental results have previously demonstrated high fidelity transportation of spin qubits between two quantum dots in silicon and identified possible sources of error. In this theoretical study, we investigate these errors and analyze the impact of tunnel coupling, magnetic field and spin-orbit effects on the spin transfer process. The interplay between these effects gives rise to double dot configurations that include regimes of enhanced decoherence that should be avoided for quantum information processing. These conclusions permit us to extrapolate previous experimental conclusions and rationalize the future design of large scale quantum processors.
\end{abstract}

\maketitle

\section{Introduction}
\label{sec:intro}

Silicon-based quantum devices are one of the brightest prospects for the future of quantum computing, achieving long coherence times \cite{pla2013high,veldhorst2015two,huang2019fidelity,yoneda2018quantum} and having potential for mass-production \cite{ansaloni2020single,zwerver2021qubits,wu2021strong}. In building a full-scale quantum computer, achieving fault-tolerance is also of utmost importance as devices are scaled up in complexity and number of qubits \cite{fowler2012surface,veldhorst2017silicon}. Scalable fault-tolerant architectures have generally been envisioned to require some form of long distance coupling between qubit arrays \cite{vandersypen2017interfacing}. Moving the qubits themselves is one of the promising strategies shown across various material platforms with both theoretical and experimental studies \cite{greentree2004coherent,bertrand2016fast,fujita2017coherent,nakajima2018coherent}. Other strategies being adopted include the coupling of spins to photons in a cavity \cite{vermersch2017quantum,mi2017strong,samkharadze2018strong,borjans2020resonant}, the use of a spin bus \cite{friesen2007efficient}, using spin chains \cite{di2010quantum}, transporting electrons with surface acoustic waves (SAW) \cite{sogawa2001transport}, and coherent SWAP gates based on exchange operations applied consecutively \cite{sigillito2019coherent}. 

Here, we focus on the coherent transfer of spins by shuttling an electron qubit between MOS quantum dots in silicon, including both analysis of experimental demonstrations \cite{hensen2020silicon,yoneda2021coherent} and theoretical studies \cite{ginzel2020spin,buonacorsi2020simulated,krzywda2020adiabatic}. It has been recently demonstrated that it is possible to perform high fidelity transport of spins in a double quantum dot with a polarization transfer fidelity of 99.97\% and average coherent transfer fidelity of 99.4\% \cite{yoneda2021coherent}. These results will serve as the motivation for the theory presented in this paper. 

One of the key challenges for coherent transport in silicon MOS devices arises from variability in the $g$-factors of each quantum dot in a multi-dot architecture, leading to different Zeeman splittings in each of the dots. This variability occurs primarily due to surface roughness at the interface between silicon and silicon dioxide in MOS devices. If there is a large Zeeman splitting difference between the dots, it would make the spin transfer process more difficult and expose the qubits to more errors. In general, there are several methods of minimizing a large difference in Zeeman splitting. One example is to do so via some engineering of either the device fabrication process or lowering the magnetic field. The scenario in which we have a large difference in Zeeman splitting is also an important one to consider especially for examining the transfer fidelity of a long chain of dots. In this paper, we investigate the transport process and the sweet spot near the interdot transition. We are still able to draw general qualitative conclusions while quantitative conclusions will be based on the case of a large Zeeman splitting difference (with its value taken from experiments done in Ref.~\cite{yoneda2021coherent}).

In the next section (Section~\ref{sec:model}), we will present the theory and modeling of the double quantum dot system Hamiltonian and the qubit dispersion. In Section~\ref{sec:noise}, we examine the temporal errors using a four-level model for the qubit, where we solve the time independent Schrödinger equation while subjected to computer-generated noise. Here, we examine the impact of noise on both the Ramsey-like coherence times and Hahn echo times. In this section, we also demonstrate the presence of a sweet spot in coherence times. Following that, in Section~\ref{sec:gates}, we reduce further into a two-level system and examine another possible source of the transfer error in the form of unwanted rotations on the Bloch sphere during the transfer process. Finally, in Section~\ref{sec:discuss}, we discuss potential improvements to the shuttling process to improve the transfer fidelity.

\section{Theory and Modeling}
\label{sec:model}

In this section, we define the Hamiltonian of our double quantum dot system and examine the qubit dispersion that results.  Typically, electrons in silicon have a total of six valley states, with two states per Cartesian direction. In silicon devices, the electric confinement of the quantum dots results in the valley states along the $x$ and $y$ directions being much higher in energy than in the $z$ direction, resulting in only two valley states relevant to our discussion \cite{kane2000silicon,ando1982electronic}. The valley states will be defined using an effective mass approach \cite{saraiva2011intervalley} where the valley wavefunctions are described by the combination of the wave envelope function in the $z$ direction and Bloch functions.

In our system, we have a single electron in a double quantum dot in silicon. We can define the wavefunction as follows,
\begin{align}
    \label{eqn:wavefn}
    \ket{\psi_{i,\pm}} = F_i(\mathbf{r}) u_\pm(\mathbf{r}) e^{\pm i k_0 z}\:,
\end{align}
where $F_i(\mathbf{r})$ is the envelope wavefunction in the $x$, $y$, and $z$ directions and describes the spatial part of the wavefunction, and $u_\pm(\mathbf{r}) e^{\pm i k_0 z}$ is the Bloch function describing the valley states, with the wave vector $\pm k_0 \hat{z}$ describing the positions in momentum-space of the conduction band minima. Valley eigenstates are generally a superposition of the $+k_0$ and $-k_0$ valley states, with their coefficients determined from the valley phases. This is accounted for in the Hamiltonian, which consists of charge, spin and valley degrees of freedom. Charge states here refer to the location of the electron, \textit{i.e.}, whether the electron is in dot A or dot B, and we define detuning to be the energy separation between charge states localized in the two dots. Taking into consideration all of these degrees of freedom, the basis states in which we construct our Hamiltonian will be given by,
\begin{align}
    \label{eqn:basis}
    \left\{\ket{A,\uparrow,-k_0},\ket{A,\downarrow,-k_0},\ket{A,\uparrow,+k_0},\ket{A,\downarrow,+k_0},\right.\nonumber\\\left.\ket{B,\uparrow,-k_0},\ket{B,\downarrow,-k_0},\ket{B,\uparrow,+k_0},\ket{B,\downarrow,+k_0}\right\}\:.
\end{align}

We define the Hamiltonian in second quantization in terms of the creation and annihilation operators, $c^{(\dagger)}_{i,\sigma,v}$, and the number operator, $\hat{n}_{i,\sigma,v}=c^\dagger_{i,\sigma,v} c^{}_{i,\sigma,v}$. The total Hamiltonian is given as,
\begin{align}
    \label{eqn:ham8}
    \hat{H} = \hat{H}_\mathrm{qd} + \hat{H}_\mathrm{Z} + \hat{H}_\mathrm{soc} + \hat{H}_\mathrm{valley} + \hat{H}_\mathrm{sv}\:,
\end{align}
where
\begin{multline}
    \hat{H}_\mathrm{qd} = \sum_{\sigma,v} \left[\frac{\varepsilon}{2} \left(\hat{n}_{\mathrm{A},\sigma,v} - \hat{n}_{\mathrm{B},\sigma,v}\right) + \right.\\ \left.\sum_{i\neq j}\frac{t_\mathrm{c}}{2} \left(c^\dagger_{i,\sigma,v}c^{}_{j,\sigma,v} + \mathrm{h.c.}\right)\right]
\end{multline}
is the quantum dot Hamiltonian and describes the detuning ($\varepsilon$) and spin- and valley-independent tunnel coupling ($t_\mathrm{c}$) between dots. In this study, to be consistent with the value of tunnel coupling found for the device used experimentally in ref. \cite{yoneda2021coherent}, we adopt a value of tunnel coupling corresponding to approximately 430 \textmu{}eV. While a large tunnel coupling (in this case larger than the Zeeman splitting at 1 T by a factor of a few) is advantageous in suppressing diabatic effects \cite{krzywda2020adiabatic}, we note that this is already near the upper bound of useful range given the typical valley splitting in Si-MOS dots \cite{yang2013spin}. The second term
\begin{align}
    \hat{H}_\mathrm{Z} = \sum_{i,v} \frac{E_{\mathrm{Z},i}}{2} \left(\hat{n}_{i,\uparrow,v} - \hat{n}_{i,\downarrow,v}\right)
\end{align}
describes the dot-dependent Zeeman splitting ($E_{\mathrm{Z},i}$). Due to spin-orbit coupling, the Zeeman splitting would also be valley-dependent. However, we can neglect its impact in this case since the electron will not occupy the excited valley states during the transfer process thanks to the large valley splitting in our systems (typically about 0.5 to 1 meV). Spin-orbit interaction effects can be described by,
\begin{multline}
    \hat{H}_\mathrm{soc} = \sum_{i,v}\frac{\eta_i\varepsilon}{2}\left(\hat{n}_{i,\uparrow,v}-\hat{n}_{i,\downarrow,v}\right) \\ + 
    \sum_{v, i \neq j}\frac{t_\mathrm{sd}}{2} \left(c^\dagger_{i,\uparrow,v}c^{}_{j,\uparrow,v} - c^\dagger_{i,\downarrow,v}c^{}_{j,\downarrow,v}\right) \\ +
    \frac{t_\mathrm{sf}}{2} \left(c^\dagger_{i,\uparrow,v}c^{}_{j,\downarrow,v} + c^\dagger_{i,\downarrow,v}c^{}_{j,\uparrow,v}\right) + \mathrm{h.c.} \:.
\end{multline}
Note that the Zeeman splitting differing in each dot is also a result of spin-orbit coupling. The first spin-orbit effect here is the linear Stark shift denoted by $\eta_i$. The second effect alters the tunneling process and includes both spin-dependent ($t_\mathrm{sd}$) and spin-flip ($t_\mathrm{sf}$) terms. Spin-dependent effects alter the tunnel coupling such that different spin states are coupled at slightly different amplitudes, whereas spin-flip effects couple the spin up and down states in different charge states. The coupling between valley states is described by,
\begin{align}
    \hat{H}_\mathrm{valley} = \sum_{i,\sigma} E_{\mathrm{v},i} e^{i\phi_i} \left(c^\dagger_{i,\sigma,v}c^{}_{i,\sigma,v^\prime}\right) + \mathrm{h.c.} \:,
\end{align}
where $E_{\mathrm{v},i}$ is the intensity of the valley coupling and $\phi_i$ is the valley phase, which differs for each dot. The final term in the Hamiltonian describes spin-valley mixing and is given by,
\begin{multline}
    \hat{H}_\mathrm{sv} = \sum_{i} \frac{\Delta^\mathrm{sv}_1}{2}(c^\dagger_{i,\downarrow,-k_0}c^{}_{i,\uparrow,+k_0}) \\ + \frac{\Delta^\mathrm{sv}_2}{2}(c^\dagger_{i,\downarrow,+k_0}c^{}_{i,\uparrow,-k_0}) + \mathrm{h.c.} \:,
\end{multline}
which results from the valley-dependent spin-orbit field created by the $\mathrm{SiO_2}$ interface, with $\Delta_1^\mathrm{sv}=|\Delta_1^\mathrm{sv}|e^{i\phi_i}$ and $\Delta_2^\mathrm{sv}=|\Delta_2^\mathrm{sv}|e^{i\phi_i}$ \cite{huang2014spin,zhang2020giant,cai2021coherent}.

\begin{figure}[ht!]
    \centering
    \includegraphics[width=0.44\textwidth]{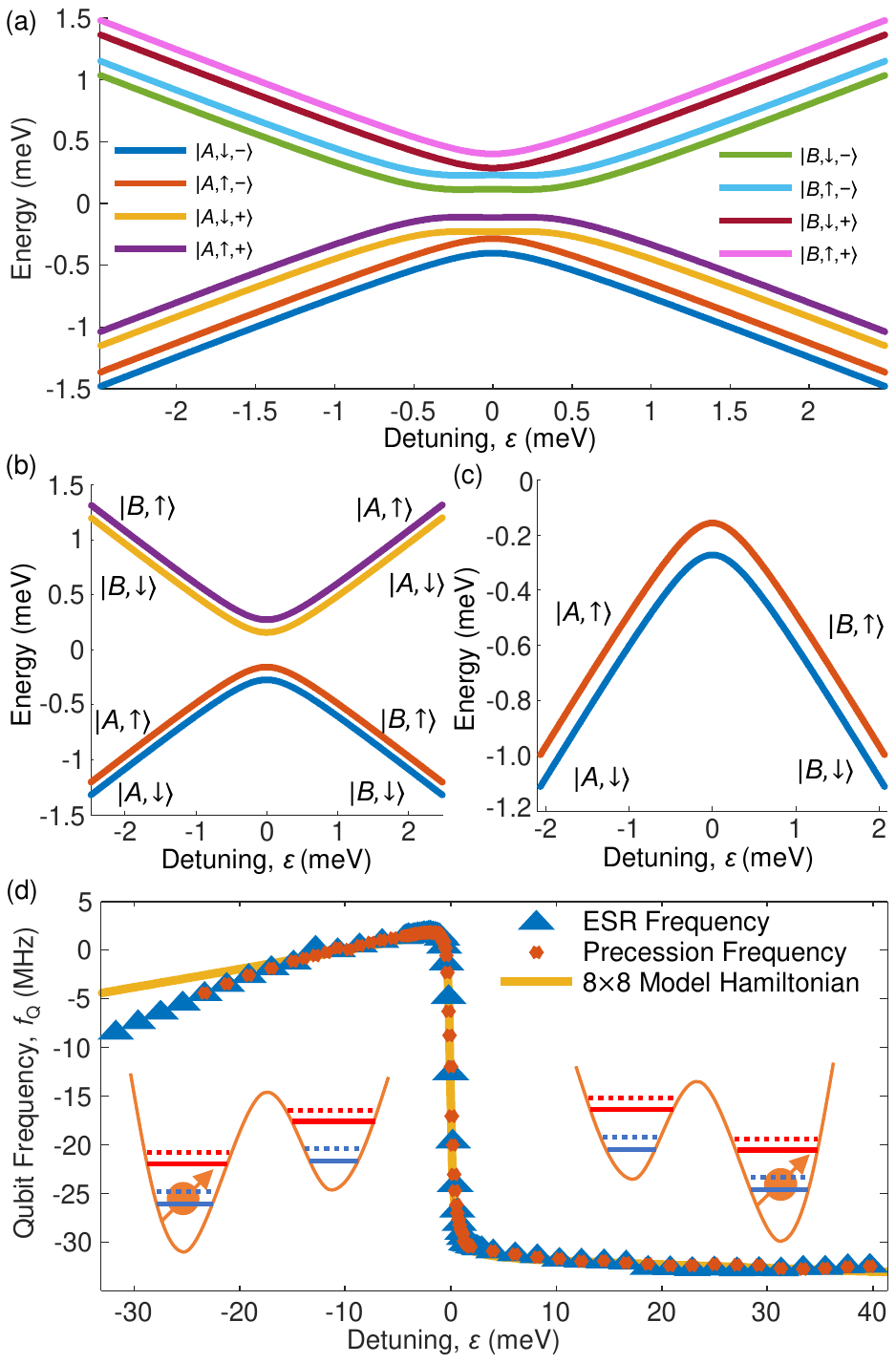}
    \caption{Energy levels and qubit dispersion in regime $t_\mathrm{c} \gg E_\mathrm{Z}$. (a) shows the energy levels of the full $8\times8$ Hamiltonian as described in Eq.~(\ref{eqn:ham8}). The legend describes the states when the electron is far detuned in dot A ($\varepsilon \ll -t_\mathrm{c}$), with $\ket{-}$ and $\ket{+}$ being the valley eigenstates. (b) shows the energies of the four-level model where we consider only the charge (electron position) and spin states. (c) shows the two-level system of only the effective ground states obtained from a Schrieffer-Wolff transformation. (d) shows the qubit dispersion of the double quantum dot system. The electron spin resonance (ESR) frequency (blue triangles) and precession frequency (red circles) are obtained experimentally as shown in ref.~\onlinecite{yoneda2021coherent}. The yellow line plots the fit of the dispersion calculated from the eight-level Hamiltonian model [Eq.~(\ref{eqn:ham8})]. Magnetic field, $B_0$, is set at $1~\mathrm{T}$ here. The accompanying schematic demonstrates the location of the electron as detuning changes. There are a total of eight energy levels with the red and blue levels respectively representing the ground and excited valley eigenstates. The dotted and solid lines are spin up and down states respectively.}
    \label{fig:qubitfreq}
\end{figure}

The Hamiltonian defined above will serve as the basis for all the calculations that follow in the rest of the paper. Depending on the situation and the type of analysis, we employ different Hamiltonians defined in different bases for ease of understanding the specifics of the system. In general, if we are to consider all eight basis states as described in Eq.~\ref{eqn:basis}, we would obtain an eight-level energy diagram as shown in Fig.~\ref{fig:qubitfreq}(a). In Section~\ref{sec:noise}, where we analyze the effect of charge noise, we consider only the Hamiltonian in the lower valley eigenstate. Therefore, the basis states considered here become $\left\{\ket{A,\uparrow},\ket{A,\downarrow},\ket{B,\uparrow},\ket{B,\downarrow}\right\}$. In Section~\ref{sec:gates}, where we are interested in unitary errors that occur as unwanted rotations, we consider the Hamiltonian in the reduced Schrieffer-Wolff basis. The Schrieffer-Wolff method is a quasi-degenerate perturbative method which projects the unwanted excited states into the desired ground states \cite{winkler2003spin}. In all of the remainder of the paper, we are performing numerical simulations assuming a specific set of parameters corresponding to that fitted from experimental data from Ref.~[\onlinecite{yoneda2021coherent}] or from literature values, which are summarized in Appendix~\ref{app:hamiltonian}. We show the energy diagrams of the respective multi-level models in Figs.~\ref{fig:qubitfreq}(a)-(c). Specifically, we plot the qubit frequency obtained based on the first excitation energy of the eight-level model in Fig.~\ref{fig:qubitfreq}(d). As a benchmark, we plot the qubit energy diagrams predicted by the different models for the same set of parameters in Fig.~\ref{appfig:qubitfreq} in Appendix~\ref{app:hamiltonian}.

The primary topic of study in this paper is the coherent transfer of spin qubits in double quantum dots, the schematic of which is shown in Fig.~\ref{fig:qubitfreq}(d). Typically, the qubit state is initialized in one of the quantum dots, for example, dot A in the case of $\varepsilon \ll -t_\mathrm{c}$, with the ground state being the superposition state of spin up and down in the lower valley eigenstate. A pulse in voltage detuning brings the energy levels of the two dots into equilibrium. Near this equilibrium point ($\varepsilon=0$), charge hybridization occurs, and the wavefunction of the electron is spread across both quantum dots. Finally, after the transfer, the qubit state is now in dot B ($\varepsilon \gg t_\mathrm{c}$).

We will test the model defined above against the experimental results obtained in Ref.~(\onlinecite{yoneda2021coherent}). We reproduce in Fig.~\ref{fig:qubitfreq}(d) the results from two experiments \cite{yoneda2021coherent}. The blue triangles are the resonant frequencies for microwave spin driving, and the red dots are the frequency of qubit precession measured from a Ramsey experiment. The precession frequency of the qubit is the frequency at which the qubit accumulates phase and will be referred to as the qubit frequency. The qubit frequency changes as the electron moves from one dot to another.

In comparison, we can model the qubit frequency using the Hamiltonian as defined in Eq.~(\ref{eqn:ham8}) while taking into account all eight energy levels as shown in Fig.~\ref{fig:qubitfreq}(a) and with basis states as defined in Eq.~(\ref{eqn:basis}). The qubit frequency can be obtained by calculating the first excitation energy, with the result shown in Fig.~\ref{fig:qubitfreq}(d) (yellow line).

This eight-level model describes the qubit dispersion except for nonlinear Stark shifts at far detuning levels. Therefore, our analysis will be limited to voltage detuning within $\varepsilon = \pm2.5~\mathrm{meV}$ and will not be impacted by nonlinear Stark shift effects in far detuning levels. This model serves as the theoretical basis upon which we build our understanding of the double dot system and the spin transport process. In recent experimental findings relating to spin transport \cite{yoneda2021coherent}, one of the key findings is that there are two main types of errors that accumulate during the transfer process. One is an error accumulated over time due to the exposure of the qubit to electric field fluctuations, which can be described by simulating Ramsey-like, $T_2^*$, and Hahn echo, $T_2^\mathrm{H}$, coherence times. The other error component is a transfer error that does not depend on the ramp time but increases with the number of transfers across the inter-dot region. In this work, we examine both of these sources of error, investigating possible causes and their impacts on the transfer process.

\section{Electric noise and sweet spots}
\label{sec:noise}

\begin{figure*}[ht]
  \centering
  \includegraphics[width=0.9\textwidth]{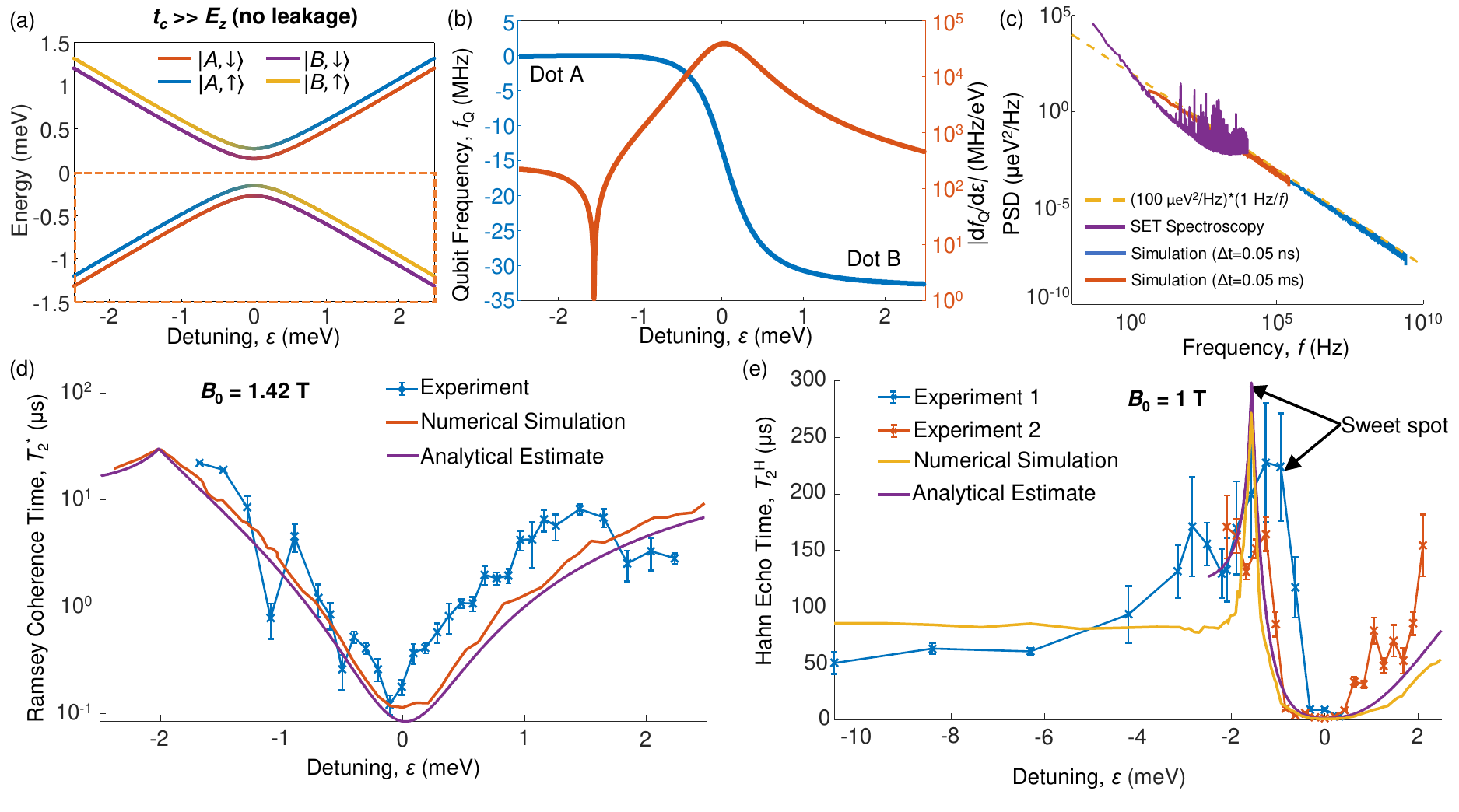}
  \caption{Effect of $1/f$ charge noise on Ramsey and Hahn echo coherence times. (a) shows the energy diagram for a four-level model with spin and charge degrees of freedom, in the regime where $t_\mathrm{c} \gg E_\mathrm{Z}$, and colors are coded based on the proportion of each state as per the legend. We will focus on the ground charge states for next section (enclosed in orange dotted lines). (b) shows the qubit dispersion ($f_\mathrm{Q}$) and its gradient ($\left|\mathrm{d}f_\mathrm{Q}/\mathrm{d}\varepsilon\right|$), defined with respect to the qubit Larmor frequency at $\varepsilon \sim -1.35~\mathrm{meV}$ (dot A). The gradient goes to zero at $\varepsilon=-1.56~\mathrm{meV}$. (c) shows the results of the single electron transistor (SET) noise spectroscopy measurement (in purple). Trend line in yellow shows the extrapolation of the noise amplitude at $1~\mathrm{Hz}$ (100 \textmu{$\mathrm{eV}^2/\mathrm{Hz}$}) to higher frequencies. Examples of computer-generated noise spectra are also plotted at sampling frequencies of 20 GHz (blue) and at 20 kHz (red). (d) shows the $T_2^*$ Ramsey coherence times for different values of detuning at $B_0=1.42~\mathrm{T}$. The sweet spot in coherence is observed at about $-2~\mathrm{meV}$ at this magnetic field. We also plot here in purple $\left|(\mathrm{d}f_\mathrm{Q}/\mathrm{d}\varepsilon)\delta\varepsilon\right|^{-1}$ calculated at $B_0=1.42~\mathrm{T}$, with $|\delta\varepsilon|\approx200~\text{\textmu{}}\mathrm{eV}$. (e) shows the $T_2^\mathrm{H}$ Hahn echo times at $B_0=1~\mathrm{T}$. We show here two sets of experimental data from a Hahn echo-like experiment performed in a device reported in Ref.~\cite{yoneda2021coherent}, with only the data from experiment 1 showing a sweet spot in detuning.}
  \label{fig:error}
\end{figure*}

In this section, we look at temporal errors, and how these errors can impact the transfer process. We focus on the variation of coherence times ($T_2^*$ and $T_2^\mathrm{H}$) with detuning, especially close to the inter-dot anticrossing.

The limitations on $T_2^*$ coherence times vary from system to system, but for our system of concern [that in Ref.~(\onlinecite{yoneda2021coherent})], the same device has been extensively characterized in a previous experiment in Ref.~(\onlinecite{chan2018assessment}). There, the two main sources of noise identified were the hyperfine coupling to the residual ${}^{29}\mathrm{Si}$ nuclear spins \cite{khaetskii2002electron} (the device is fabricated on isotopically purified ${}^{28}\mathrm{Si}$, with residual ${}^{29}\mathrm{Si}$ at concentrations of 800ppm) and the $1/f$ noise originating from fluctuations of two-level charged systems in the oxide and at the interface between the oxide and silicon \cite{culcer2009dephasing,kuhlmann2013charge}, which create noise on the spin mediated by the spin-orbit coupling. For this analysis, we will focus on the $1/f$ noise effects, which becomes dominant with the strong dispersion at the onset of the dot transition. Hyperfine effects are also expected to have some influence on the qubit shuttling performance when the variability between qubit frequencies in the dots is not too large, but for the regime considered here we disregard such effects.

For the purpose of understanding the coherence times, we neglect the valley degree of freedom, which was defined in Eq.~(\ref{eqn:ham8}), leaving us with an effective four-level system. In our system, the valley splittings are expected to be on the order of 0.5 to 1 meV and therefore it is unlikely for the electron to couple to these valley states during the transport process. In this case, the charge and spin states will be sufficient for an analysis of the impact of $1/f$ noise on the system, with charge noise entering directly through the charge degree of freedom, degrading the spin coherence through the dependence of its frequency on $\varepsilon$. Note that valleys still play a role in this system because the linear spin-orbit coupling of each dot is dictated by its valley structure \cite{ruskov2018electron}.

We can observe in Fig.~\ref{fig:error}(a) the energy levels relevant to our analysis. At negative detuning values, the ground state wavefunctions of the system are primarily in dot A, and correspondingly, the ground state wavefunctions are mostly in dot B at positive detuning values. There is a large energy gap between the two lowest levels and the excited pair of states, corresponding to the large tunnel coupling in our double dot system ($\sim 430~\text{\textmu{}}\mathrm{eV}\approx h\times104~\mathrm{GHz}$). In this regime, the tunnel coupling is much larger than the Zeeman splitting at 1T, and this reduces the state leakage in the transfer process.

In Fig.~\ref{fig:error}(b), we plot the dispersion of the first excitation energy of the double dot system. The difference in Zeeman splittings between the two dots is caused by the surface roughness arising from atomic sources of disorder in the oxide as mentioned before. This creates a small difference in the $g$-factors of the two dots that generally results in tens of megahertz of difference in the Zeeman splittings \cite{tanttu2019controlling}. The transition between these two qubit frequencies is set by the charge hybridization between the dots due to the tunnel coupling.

Examining the qubit frequency, we find that it is highly dependent on the detuning energies, especially near the anticrossing, where there is a steep transition from one qubit frequency to another. Charge noise causes fluctuations of the dot levels and therefore enters as detuning noise in the Hamiltonian. Near the anticrossing at zero detuning, small fluctuations in detuning leads to large shifts in frequency, due to the large Stark shifts, $\left|\mathrm{d}f_\mathrm{Q}/\mathrm{d}\varepsilon\right|$, plotted (in red) in Fig.~\ref{fig:error}(b).

In reality, $1/f$ noise is not always small in amplitude, such that analyzing $\left|\mathrm{d}f_\mathrm{Q}/\mathrm{d}\varepsilon\right|$ may be insufficient in some scenarios. To understand how the coherence times are related to the qubit frequency, we simulate the effect of charge noise on the qubit. The Hamiltonian we use to describe our system is given explicitly as,
\begin{widetext}
\begin{align}
    \hat{H}_{4\times 4} = \frac{1}{2}\begin{pmatrix}
    E_\mathrm{Z,A} + \left(\eta_\mathrm{A} + 1\right)\varepsilon & 0 & t_\mathrm{c} + t_\mathrm{sd} & t_\mathrm{sf} \\
    0 & -E_\mathrm{Z,A} - \left(\eta_\mathrm{A} - 1\right)\varepsilon & t_\mathrm{sf} & t_\mathrm{c} - t_\mathrm{sd} \\
    t_\mathrm{c} + t_\mathrm{sd} & t_\mathrm{sf} & E_\mathrm{Z,B} + \left(\eta_\mathrm{B} - 1\right)\varepsilon & 0 \\
    t_\mathrm{sf} & t_\mathrm{c} - t_\mathrm{sd} & 0 & -E_\mathrm{Z,B} - \left(\eta_\mathrm{B} + 1\right)\varepsilon
    \end{pmatrix}\:,
    \label{eqn:ham4x4}
\end{align}
\end{widetext}
where all the terms are as defined in the previous section. We can model $1/f$ charge noise in this system as fluctuations on the detuning levels. Thus, the noise Hamiltonian is given by,
\begin{widetext}
\begin{align}
    \hat{H}_\mathrm{noise}(t) = \frac{1}{2}\begin{pmatrix}
     \left(\eta_\mathrm{A}+1\right)\delta\varepsilon(t) & 0 & 0 & 0 \\
    0 &  -\left(\eta_\mathrm{A}-1\right)\delta\varepsilon(t) & 0 & 0 \\
    0 & 0 & \left(\eta_\mathrm{B}-1\right)\delta\varepsilon(t) & 0 \\
    0 & 0 & 0 & -\left(\eta_\mathrm{B}+1\right)\delta\varepsilon(t)
    \end{pmatrix}\:,
    \label{eqn:Hnoise}
\end{align}
\end{widetext}
where the $\delta\varepsilon(t)$ terms are numerically generated noise in the time domain. Examples of the spectrum are shown in Fig.~\ref{fig:error}(c) and details of how the noise is simulated can be found in Appendix~\ref{app:1fsimulation}. The total Hamiltonian will then be given as,
\begin{align}
    \label{eqn:ham4}
    \hat{H}_\mathrm{total}(t) = \hat{H}_{4\times 4} + \hat{H}_\mathrm{noise}(t)\:.
\end{align}
Having defined the Hamiltonian, we now evaluate the evolution of the wavefunctions under the Hamiltonian by solving numerically the Schrödinger equation,
\begin{align}
\label{eqn:schrodinger}
    i\hbar\frac{\partial}{\partial t}\psi(t) = \hat{H}_\mathrm{total}\psi(t)\:.
\end{align}
In general, the solution to this system can be constructed from small time steps $\delta t$ as,
\begin{align}
\label{eqn:timeevo}
    \psi(t+\delta t) = e^{-i\hat{H}_\mathrm{total}\delta t/\hbar} \psi(t) \:,
\end{align}
where the time evolution of the wavefunction is governed by the unitary, $U=e^{-i\hat{H}_\mathrm{total}\delta t/\hbar}$. This approximation is valid as long as $\delta t$ is much smaller than the characteristic time scale of variation of $\hat{H}_\mathrm{total}$. We initialize at $t=0$ into the superposition state of spin up and down, and then iteratively calculate the unitary and the wavefunction of Eq.~(\ref{eqn:timeevo}) until a total evolution time $t_\mathrm{evol}$.

We note that only the time dependence of the noise Hamiltonian is kept in the total expression for the Hamiltonian, with the qubit Hamiltonian time-independent in Eq.~(\ref{eqn:ham4}). In general, the qubit Hamiltonian ($\hat{H}_{4\times4}$) is also time-dependent since it contains the detuning parameter $\varepsilon(t)$. This implies that in order for the approximation in Eq.~(\ref{eqn:timeevo}) to be valid, we would have to use small time discretization of at least three orders smaller than the Hamiltonian terms for our purpose since the detuning is the dominating energy term. This is also the case for the calculations in the next section where we do not consider any noise in the system and examine the impact of diabatic effects on the transfer process.

However, in the case relevant to the discussion in this section, we are calculating the coherence times of the qubit for fixed detuning values. This leaves only the time dependence in the noise Hamiltonian. We do not have to sweep the detuning values over time. The only time-varying parameters are the noise parameters which are themselves small. This allows us to perform the simulation using coarse time steps of a tenth of the total evolution time, $t_\mathrm{evol}$/10, which, as we will show later in the section, reproduces the experimental results sufficiently well.

A computational challenge is the fact that the calculation has to be performed over several orders of magnitude in time. There is a large range of expected coherence times, with minimum coherence times on the order of 0.1 \textmu{}s near the anticrossing, and maximum coherence times near the sweet spot expected to be on the order of hundreds of microseconds. In order to capture the full range of coherence decay, we divide up the numerical simulation into different sections with each section consisting of $t_\mathrm{evol}$ varying over a single order of magnitude, thus speeding up significantly the simulation. More details are contained in Appendix~\ref{app:numerics}.

Our simulated noise is calibrated against the noise amplitude measured from the current through a single electron transistor (SET) near the quantum dots, which is used to estimate the electric noise in the device (Appendix~\ref{app:setspec}).

The purple trace in Fig.~\ref{fig:error}(c) shows the results from this experimental technique, and we take reference to its amplitude at $f=1~\mathrm{Hz}$, with the yellow dotted line as the reference line for $1/f$ noise of amplitude $100~\text{\textmu{}}\mathrm{eV}^2/\mathrm{Hz}$. The power spectral density (PSD) for two examples of the computer generated noise is also plotted in Fig.~\ref{fig:error}(c) corresponding to different sampling frequencies ($20~\mathrm{GHz}$ for the blue trace and $20~\mathrm{kHz}$ for the red trace).

Given the statistical nature of the noise, we repeat the calculation of the density matrix $\rho$ after $t_\mathrm{evol}$ over 100 realizations of noise. The final density matrix is then averaged over these 100 iterations and we obtain $\bar{\rho}$. Finally, we calculate the Bloch length \cite{kimura2003bloch,jakobczyk2001geometry}, which is a measure of the qubit coherence,
\begin{align}
    \left|\mathbf{r}\right| = \sqrt{\frac{4}{3}\left(\mathrm{Tr}(\bar{\rho}^2)-\frac{1}{4}\right)} \:.
\end{align}
The Bloch length is chosen here because it corresponds to what was measured in the experiment in Ref.~(\onlinecite{yoneda2021coherent}). We can calculate the Bloch length as a function of time evolution, $t_\mathrm{evol}$. We obtain a decay, which can be fitted using, $|\mathbf{r}| = A\exp(-(t_\mathrm{evol}/T_2^*)^\beta) + C$, where $A$, $C$, $T_2^*$, and $\beta$ are the amplitude, the final Bloch length after decay, the coherence time, and the decay exponent, respectively. The same expression is adapted later to obtain the Hahn echo time.

We used these numerical simulations to estimate the effect of $1/f$ charge noise on the system. In addition, we also account for hyperfine nuclear spin noise $1/T_2^\text{hyp}$ \cite{hensen2020silicon}, which combine with the electric charge noise to determine the $T_2^*$ coherence time,
\begin{align}
    \frac{1}{T_2^\mathrm{*}} = \frac{1}{T_2^\mathrm{elec}} + \frac{1}{T_2^\mathrm{hyp}} \; \text{,}
\end{align}
Hyperfine nuclear noise may be relevant when we suppress the effect of charge noise at the sweet spot and this source of noise is modeled to be independent of detuning and would set an upper bound for the coherence time. We estimate the amplitude $T_2^\text{hyp}$ to be approximately $30~\text{\textmu{}s}$.

Following this process, we obtain the results shown in Fig.~\ref{fig:error}(d). Here, we set the $B_0$ magnetic field to $1.42~\mathrm{T}$, corresponding directly to the conditions under which the experiment in Ref.~(\onlinecite{yoneda2021coherent}) was performed, which yielded the experimental data (plotted in blue) in Fig.~\ref{fig:error}(d). We plot the Ramsey coherence times to highlight the point of minimum $T_2^*$ near the anticrossing at zero detuning, and we also obtain fairly consistent results between the numerical simulation (in red) and the experiment (in blue). 

In addition to the numerical simulations, we also performed an analytical estimate of the electric charge noise, where we approximate its effect on the spin to be proportional to its Stark shift and therefore, we can describe the decoherence due to charge noise in the following way,
\begin{align}
    \frac{1}{T_2^\mathrm{elec}} = \left|\frac{df_\mathrm{Q}}{d\varepsilon}\right||\delta\varepsilon|
\end{align}
where $\left|\frac{df_\mathrm{Q}}{d\varepsilon}\right|$ can be obtained directly from Fig.~\ref{fig:error}(b) while $|\delta\varepsilon|$ can be estimated from the amplitude of $1/f$ noise at 1 Hz using the following expression $\left(\sim 2\pi |\delta\varepsilon|_\mathrm{1Hz}\sqrt{\ln{\left(\frac{f_\mathrm{h}}{f_\mathrm{l}}\right)}}\right)$ \cite{yoneda2018quantum,cywinski2008enhance}. We estimate $|\delta \varepsilon|_\text{1Hz}$ to be $10~\text{\textmu{}eV}$ from the amplitude spectral density of  $10~\text{\textmu{}eV}/\sqrt{\text{Hz}}$ at 1Hz in our charge noise model. Based on the way we generate the noise spectra, we estimate our ratio of $f_\text{h}/f_\text{l}$ to be approximately $5\times10^4$, given that $f_\text{h}=1/(2\Delta{t})$ and $f_\text{l}=1/(N\Delta{t})$ (Appendix~\ref{app:1fsimulation}). This leads to a value of $\sim200~\text{\textmu{}eV}$ for $|\delta\varepsilon|$. As with the numerical simulations, we estimate the $T_2^\text{hyp}$ to be approximately $30~\text{\textmu{}s}$. We can observe that the analytical estimation of the $T_2^\text{*}$ coherence time spectrum is also fairly accurate with both the detuning dependence modeled accurately as well as the expected trend at the sweet spot based on the experimental data.

We find that there is good agreement between the experimental results with both the full numerical simulation and the analytical estimation, suggesting that outside of the sweet spot, most of the noise in our system is quasi-static charge noise in this interdot tunneling regime.

We observe that there is indeed a dip in the coherence time near zero detuning, where the inter-dot anticrossing is. This coincides with what we observed from the Stark shift of the double quantum dot [as shown in Fig.~\ref{fig:error}(b)] where it is maximum close to the inter-dot anticrossing. We also observe a coherence time sweet spot from the simulation results which is outside of the range of detuning values swept in the experiment. We note that the point of maximum coherence time is indeed at the point where $df_\mathrm{Q}/d\varepsilon=0$. 

Next, we incorporate a Hahn echo pulse into the simulation. The key difference between this simulation and what was described previously is the inclusion of a $\pi$-pulse during the time evolution of the qubit state. The $\pi$-pulse is implemented by applying a unitary operator,
\begin{align}
    U_\pi = \exp(-i~\frac{\pi}{2}~\tau_I \otimes \sigma_y) \:,
    \label{eqn:ypulse}
\end{align}
with
\begin{align}
    \tau_I \otimes \sigma_y = \begin{pmatrix}
    0 & -1i & 0 & 0 \\ 1i & 0 & 0 & 0 \\ 0 & 0 & 0 & -1i \\ 0 & 0 & 1i & 0
    \end{pmatrix} \:,
\end{align}
where $\tau$ and $\sigma$ are the Pauli matrices representing the charge and spin subspace respectively, and $U_\pi$ represents a $\pi$-pulse about the $y$-axis in the spin-subspace. This unitary operator is multiplied to the qubit wavefunction at exactly the mid-point of the time evolution ($t_\mathrm{evol}/2$), corresponding also to the pulsing sequence used in the experiment \cite{yoneda2021coherent}. Otherwise, the other aspects of calculating the Hahn echo time, $T_2^\mathrm{H}$, are the same as for the Ramsey coherence time, $T_2^*$.

The results of this simulation are shown in Fig.~\ref{fig:error}(e), where we can also observe both the point of minimum Hahn echo times near the inter-dot anticrossing and the sweet spot in detuning. At the sweet spot, we benefit from enhanced coherence, which is indicative of an ideal point for qubit idling due to reduced temporal errors. The existence of this sweet spot can be explained by examining the qubit dispersion plotted in Fig.~\ref{fig:error}(b) and the plotted gradient $df_\mathrm{Q}/d\varepsilon$, where we observe that the dispersion gradient vanishes at $\varepsilon=-1.56~\mathrm{meV}$. In the context of qubit transfer, this is an ideal point in detuning for initializing or idling due to its long coherence times. Such a sweet spot could potentially also be an ideal point for qubit control.

Similar to the case of $T_2^*$, we also plot here in purple in Fig.~\ref{fig:error}(e) the theoretical estimation of the Hahn echo time $T_2^\text{hyp}$  as well as include the effect of hyperfine nuclear noise in both the numerical simulation and analytical estimations. Since the Hahn echo is the result of a single decoupling pulse, in the specific case of $1/f$ noise, previous studies have shown that the average noise amplitude across all frequencies would be reduced by approximately an order of magnitude \cite{cywinski2008enhance,nakamura2002charge,ithier2005decoherence}. Therefore, we estimate the reduced charge noise after the Hahn echo to be $|\delta\varepsilon|\sim20~\text{\textmu{}eV}$. Similarly, we estimate that the effect of hyperfine noise will also be reduced by an order of magnitude such that now, $1/T_2^\text{hyp}=1/(300~\text{\textmu{}s})$.

Comparing the experimental result of Fig.~\ref{fig:error}(e) in blue (experiment 1) with the simulation results, we observe a small constant offset between the results, which is discussed in Appendix~\ref{app:sweetspot}. We also discuss why in some data, such as shown in red (experiment 2), the sweet spot seems absent.

Finally, comparing $T_2^*$ and $T_2^\mathrm{H}$, we observe that the Hahn echo improves the coherence times across all detuning values up to an order of magnitude. We also note that while these results are calculated at different magnetic fields, we find that the difference in magnetic field only leads to marginal change in the results, as discussed in Appendix~\ref{app:noise}, indicating that the improvement occurs due to the echo decoupling.

Overall, there is satisfactory consistency between the experimental and simulation results for both cases of $T_2^*$ and $T_2^\mathrm{H}$, suggesting that $1/f$ noise can be used to model decoherence in our system. What these results mean in the context of qubit transport is that we should avoid idling at the regions of reduced coherence times by implementing fast pulsing between the quantum dots while making use of the sweet spot in detuning as an idling point in experiments. However, fast pulsing can also increase leakage errors, so the pulse speed would be limited by the size of the tunnel coupling \cite{krzywda2020adiabatic}. We will explore the impact of pulsing speed in more detail in Sec.~\ref{sec:discuss}.

\section{Transport process as a single qubit gate}
\label{sec:gates}

Other than the errors due to $1/f$ charge noise investigated in the previous section, we are also interested in other errors that occur near the anticrossing. Recent studies suggest that $1/f$ noise can also lead to diabatic effects at the inter-dot transition \cite{krzywda2020adiabatic}. Previous experimental results also show error that occurs regardless of the transfer ramp times and are finite even in the idealized limit of infinitely fast transfers \cite{yoneda2021coherent}. One potential type of such errors is unitary error, \textit{i.e.}, unaccounted rotations occurring on the qubit Bloch sphere, and we will analyze them in this section as effective $x$ and $z$ gates after the transfer process.


One characteristic of our double dot system is that we have a large tunnel coupling as shown in Fig.~\ref{fig:error}(a), and that has been instrumental in allowing us to pulse qubits across $4~\mathrm{meV}$ in voltage detuning over 8 ns, without significant contributions from diabatic errors. The opposite scenario with a small tunnel coupling is discussed in Appendix~\ref{app:adiabaticity}, where we have the energy diagram of a four-level system with a tunnel coupling of $41~\text{\textmu{}}\mathrm{eV}$ or $10~\mathrm{GHz}$, which is smaller than the Zeeman splitting at $B_0=1~\mathrm{T}$ ($115~\text{\textmu{}}\mathrm{eV}$ or $28~\mathrm{GHz}$). With these Hamiltonian parameters, states with different spin and charge configurations cross around the inter-dot charge transition, $\varepsilon=0$, leading to enhanced spin-flip tunneling and larger possibilities of state leakage in the transport process. This is the regime that should be avoided in qubit transport protocols.

\begin{figure}[!ht]
  \centering
  \includegraphics[width=0.45\textwidth]{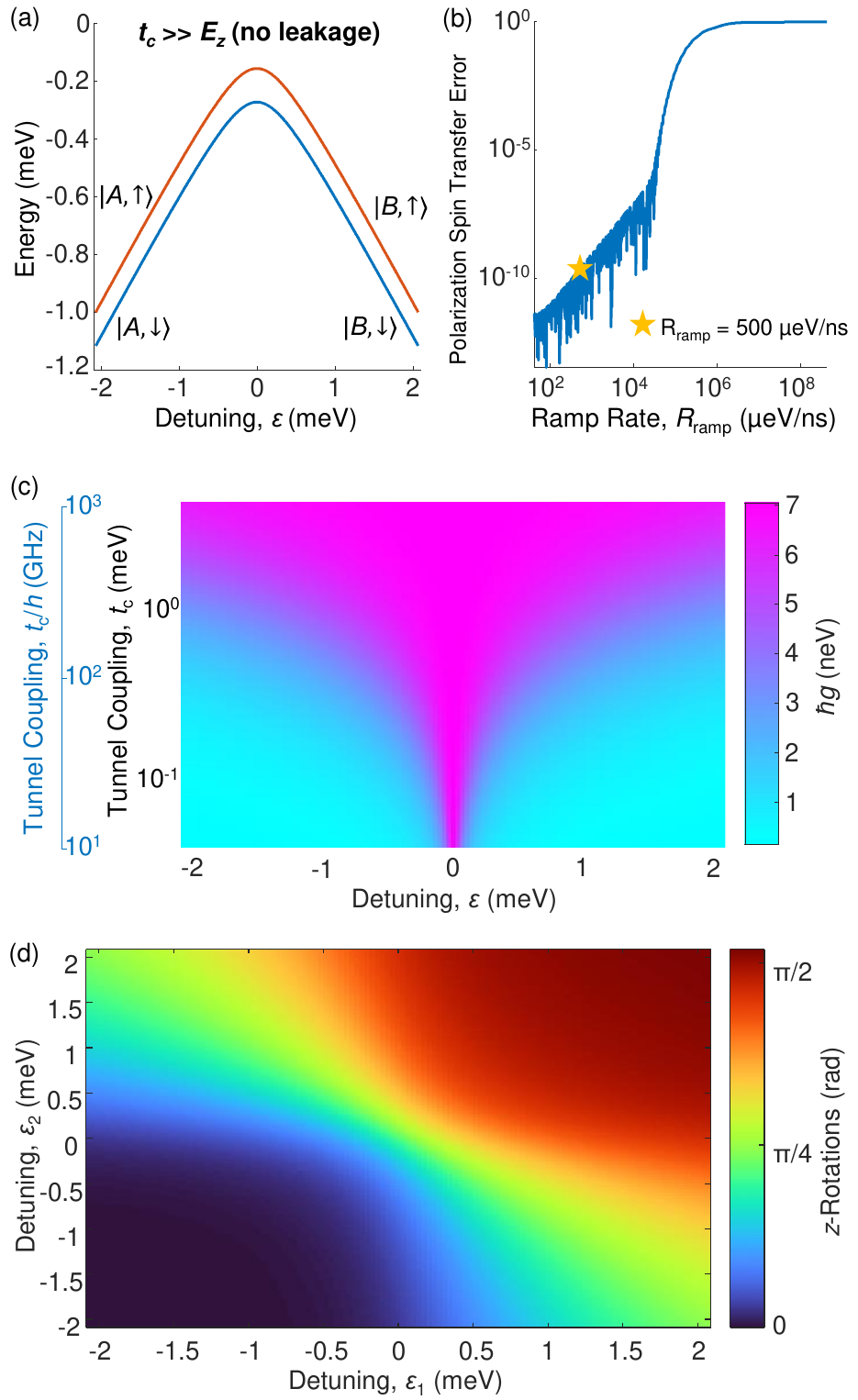}
  \caption{Unitary errors arising from the transport process. (a) shows the energy levels of the effective two-level Hamiltonian, obtained from a Schrieffer-Wolff transformation of the four-level model shown in Fig.~\ref{fig:error}(a), where $t_\mathrm{c} \gg E_\mathrm{z}$. (b) shows the polarization spin transfer errors over a range of ramp rates. The ramp rate of $500~\text{\textmu{}}\mathrm{eV/ns}$ corresponds to what is used in the experiment \cite{yoneda2021coherent} and is represented by the yellow star. (c) shows the magnitude of the off-diagonal term in the $H_\mathrm{2\times2}$ Hamiltonian. We show the magnitude of tunnel coupling in both units of meV and GHz. (d) shows the effective $z$-gate after a single transfer, which is a rotation about the quantization axis set by the $B_0$ magnetic field. We fix here the total ramp time from $\varepsilon_1$ to $\varepsilon_2$ at $8~\mathrm{ns}$. The rotating frame is defined by the precession rate at $\varepsilon=-1.35~\mathrm{meV}$ (dot A).}
  \label{fig:gates}
\end{figure}

Considering only the regime with tunnel coupling larger than the Zeeman splitting, $t_\mathrm{c}\gg E_\mathrm{z}$, we perform a Schrieffer-Wolff transformation \cite{winkler2003spin} to isolate only the effective ground state orbitals, consisting of the lowest two energy levels as shown in Fig.~\ref{fig:gates}(a). To confirm the validity of this approximation, we also calculate the spin polarization errors after a transfer across $4~\mathrm{meV}$ with ramp times spanning several different orders of magnitude, which we plot in Fig.~\ref{fig:gates}(b). For this simulation, we remain in the four-level model [Fig.~\ref{fig:error}(a)] and initialize the qubit in either the spin up or spin down state in quantum dot A, at a detuning level of $\varepsilon=-2~\mathrm{meV}$, and calculate the time evolution operators of the time-dependent Hamiltonian as we ramp the qubit from one dot to another, similar to what we outlined in Section~\ref{sec:noise}. As previously mentioned, we remove the impact of noise for this analysis and consider only diabatic effects due to the ramp itself (in a recent study it was found that the $1/f$ noise is also responsible for significant diabatic effects \cite{krzywda2020adiabatic}). 

Next, we consider a time-dependent Hamiltonian in which the detuning parameter, $\varepsilon(t)$ varies with time and $\hat{H}_\mathrm{noise}=0$. Therefore, we solve for the qubit state in time steps that are $1/10^5$ of the total ramp time, ensuring that the numerical time steps are small enough to capture the change in wavefunction accurately \cite{buonacorsi2020simulated}. Finally, we calculate the state fidelity, $\mathcal{F}$, at the end of the transfer by comparing with the target eigenstate,
\begin{align}
    \mathcal{F} = |\braket{\psi_\mathrm{target}}{\psi_\mathrm{final}}|^2 \:.
    \label{eqn:statefid}
\end{align}
We show that for a tunnel coupling of $430~\text{\textmu{}}\mathrm{eV}=h\times104~\mathrm{GHz}$, we have a very small diabatic error for our chosen ramp rate, represented by the yellow star ($500~\text{\textmu{}}\mathrm{eV/ns}$).

Now, using the Schrieffer-Wolff transformation, we derive an effective $2 \times 2$ Hamiltonian from the four-level model, with details given in Appendix~\ref{app:hamiltonian},
\begin{align}
    \hat{H}_\mathrm{2\times 2} = \frac{1}{2}\hbar\omega_0\mathbb{1} +  \frac{1}{2}\hbar\omega\sigma_z + \frac{1}{2}\hbar g \sigma_x
    \label{eqn:heff}
\end{align}
where $\omega_0$ is a shift in energy governed by the size of the tunnel coupling and the detuning, $\omega$ is dominated by the qubit frequencies of each dot depending on the detuning position and $g$ is dominated by the term $t_\mathrm{sf} t_\mathrm{c} / (2\sqrt{t_\mathrm{c}^2+\varepsilon^2)}$ which is highly dependent on the magnitude of the spin-flip tunnel coupling. We neglect second order terms and higher in  this Hamiltonian. In this form, the Hamiltonian explicitly shows that we can characterize the spin-flip error as an $x$-rotation set by $g$, and any phase accumulation as a $z$-rotation set by $\omega$.

With the effective $2\times 2$ Hamiltonian [Eq.~(\ref{eqn:heff})], we seek to understand the amount of $z$- and $x$-rotations accumulated during the transfer process in a single transfer. We will use a ramp time of 8 ns, corresponding to the ramp time used in the experiment of Ref.~(\onlinecite{yoneda2021coherent}). For these simulations, we initialized the qubit in a superposition of the two eigenstates at a particular detuning position, $\varepsilon_1$. By iteratively calculating the unitary defined in Eq.~(\ref{eqn:timeevo}), we take the product of all unitaries as we vary the detuning from $\varepsilon_1$ to $\varepsilon_2$. This gives an effective unitary time-evolution operator, $U_\mathrm{eff}$. From this effective unitary operator, we factor out a global phase $e^{i\phi}$, and then we take the matrix logarithm \cite{hansen2021pulse}. In this way, we obtain a effective operator with $\sigma_x$, $\sigma_y$, and $\sigma_z$ components, which can be interpreted as an effective qubit rotation accumulated during the transfer,
\begin{align}
    -i\log{U_\mathrm{eff}} \equiv c_1 \sigma_x + c_2 \sigma_y + c_3 \sigma_z \:,
\end{align}
where the coefficients of $c_1$, $c_2$, and $c_3$ can be obtained from this final effective operator. These coefficients are different from the terms in the 2-by-2 Hamiltonian [Eq.~(\ref{eqn:heff})]. We vary the detuning positions of the start ($\varepsilon_1$) and end ($\varepsilon_2$) points for the transfer and calculate the effective qubit rotation as a function of both these quantities.

We note that in the form of Eq.~(\ref{eqn:heff}), the leading coefficient of $\sigma_x$ is very small. In Fig.~\ref{fig:gates}(c), we show how $\hbar g$ is expected to change with both detuning $\varepsilon$ and tunnel coupling $t_\mathrm{c}$ parameters. This term is strongest near the inter-dot anticrossing, and the detuning regime in which it is significant increases with tunnel coupling. This particular form of error would benefit from a smaller tunnel coupling, but this would increase the temporal errors discussed before. Instead, faster pulsing to avoid the inter-dot region can help to minimize this error as well as the temporal errors.

Calculating the effective $x$ gates using the method above, we find that indeed the amount of $x$-rotations increases with increasing amount of time spent near the inter-dot anticrossing. However, we find that the effective $x$-rotations are small and on the order of pico-radians. This is primarily due to two reasons: the first of which is that $g$ is small compared to the $\omega$ term. Second, we define the qubit in the basis of the rotating frame, where the off-diagonal terms gain an oscillatory phase at the frequency of the rotating frame, and therefore to first order, average to zero. We estimated higher order terms in rotating wave approximation \cite{zeuch2020exact}, and that yields effective $x$-rotations on the order of pico-radians, shown in Appendix~\ref{app:xrot}. The small magnitude of the $x$-rotations is indicative that most likely, these effective rotations are not a significant source of transfer errors during the qubit transport process.

In Fig.~\ref{fig:gates}(d), we show the result of the effective $z$-rotations by plotting the coefficient $c_3$ as a function of $\varepsilon_1$ and $\varepsilon_2$. The results show that in the top right and bottom left quadrants, the amount of phase accumulated is approximately constant for a fixed total time anywhere in the quadrant. This is consistent with the fact that if the qubit is moving within the same dot and without moving too close to the anticrossing, the rate of phase accumulation is approximately constant, corresponding to the qubit frequencies of each dot in the rotating frame. The rotating frame is defined with respect to dot A, and therefore the phase accumulation in dot A is completely governed by the Stark shift in dot A, while in dot B, it will be dominated by the difference in Zeeman splitting, $\Delta E_\mathrm{Z}$. The rate of phase accumulation is then most impacted when a transfer takes place from one dot to another (\textit{i.e.} $\varepsilon_1$ and $\varepsilon_2$ not being in the same dot) or when the transfer occurs near the inter-dot anticrossing where the electron wavefunction is effectively spread across both quantum dots. 

The method described in this section was used to characterize the transfer process as an effective gate, and we found that there are unitary errors that occur as a function of detuning positions for a fixed total time. This includes a non-zero transfer error in the form of unwanted $x$-rotations, it is also too small for the particular experimental conditions to be significantly measured due to the qubit being in the rotating frame and also the small magnitude of the spin-flip term in the Hamiltonian. We expect, however, that when the Zeeman splitting matches or exceeds the tunnel coupling, this approximation will no longer hold, and significant spin flip may be observed.

\section{Discussion}
\label{sec:discuss}

In the previous sections, we have explored different error mechanisms relevant to the transport process. In this section, we discuss how this information can be of aid in future experiments in coherent transport.

\begin{figure}[!ht]
  \centering
  \includegraphics[width=0.45\textwidth]{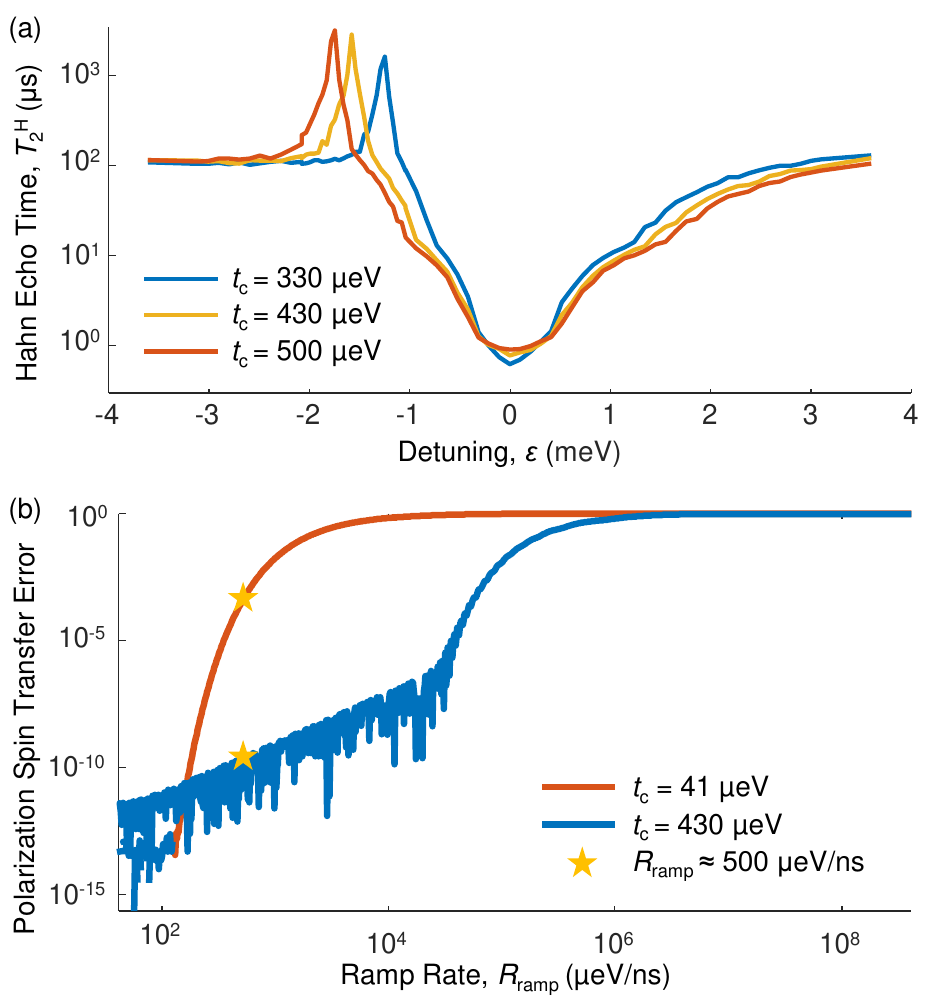}
  \caption{Impact of tunnel coupling on Hahn echo coherence times and polarization spin transfer error. (a) shows the effect of tunnel coupling on the $T_2^\mathrm{H}$ spectrum across detuning values between $-3.6~\mathrm{meV}$ and $3.6~\mathrm{meV}$. This simulation is performed at $B_0=1~\mathrm{T}$. (b) considers the transfer of a spin down ground state from dot A to B across $4~\mathrm{meV}$ with varying ramp times. The two curves are calculated for different tunnel couplings, with the red and blue traces calculated for tunnel couplings of $41~\text{\textmu{}}\mathrm{eV}\approx h\times10~\mathrm{GHz}$ and $430~\text{\textmu{}}\mathrm{eV}\approx h\times104~\mathrm{GHz}$ respectively. The yellow star indicates the ramp rate used in the experiments of Ref.~\cite{yoneda2021coherent}.}
  \label{fig:control}
\end{figure}

In the results shown in the previous sections, we set the magnitude of the tunnel coupling to be about $430~\text{\textmu{}}\mathrm{eV}\approx h\times104~\mathrm{GHz}$, obtained from experimental fits \cite{yoneda2021coherent}. Its large magnitude has been instrumental in avoiding leakage in the experiment. In this section, we examine how a different tunnel coupling can impact the results we obtained in the earlier sections.

In Fig.~\ref{fig:control}(a), we show how the Hahn echo times, $T_2^\mathrm{H}$, change with the magnitude of tunnel coupling. In Sec.~\ref{sec:noise}, we observed that coherent errors peaked at the inter-dot anticrossing and that there exists a sweet spot in detuning where the coherence time is maximum. In this figure [Fig.~\ref{fig:control}(a)], we observe that with a larger tunnel coupling, the minimum $T_2^\mathrm{H}$ time at the inter-dot anticrossing is increased slightly. Even though it remains within the same magnitude, it does suggest that a larger tunnel coupling can be helpful in extending coherence times. We also observe that the sweet spot remains visible, indicating that the sweet spot can be readily accessed in different tunnel coupling regimes and is a general characteristic of the qubit dispersion. This opens up possibilities of quantum information protocols making use of the sweet spot while having tunability of the tunnel coupling.

Next, we show in Fig.~\ref{fig:control}(b) the error rates during transfer as a function of the ramp rates, considering only diabatic effects. Here we investigate diabatic effects that occur as a result of ramping too quickly between the dots \cite{buonacorsi2020simulated}, where we calculate the state fidelity [Eq.~(\ref{eqn:statefid})] after a single transfer for different ramp rates. Similarly to Fig.~\ref{fig:gates}(b), we show the results of transferring a spin down state from dot A to B across $4~\mathrm{meV}$ with varying ramp times. In other words, this is a measurement of adiabaticity and ideally, we want minimum error rates here at the chosen ramp rate, thus minimizing the effect of state leakage. We had already seen that with a large tunnel coupling of $t_\mathrm{c}=430~\text{\textmu{}}\mathrm{eV}\approx h\times104~\mathrm{GHz}$, the error rates are very small ($\sim 10^{-10}$) at the ramp rate of $500~\text{\textmu{}}\mathrm{eV/ns}$ used in the experiment of Ref.~(\onlinecite{yoneda2021coherent}). Here, we expanded on that by also showing the results with a much reduced tunnel coupling of $t_\mathrm{c}=41~$\textmu{}$\mathrm{eV} \approx h\times10~\mathrm{GHz}$. It becomes obvious that a large tunnel coupling is more advantageous for avoiding diabatic errors by comparing the results for $\sim430~\text{\textmu{}}\mathrm{eV}$ (shown in blue) and for $\sim41~\text{\textmu{}}\mathrm{eV}$ (shown in red). The diabatic errors with a lower tunnel coupling are increased by several orders of magnitude at the same ramp rate. It would not be preferable to lower the ramp rate since one should ramp quickly across the inter-dot anticrossing in order to minimize the temporal errors. This points to a large tunnel coupling being a key step towards optimizing the transfer process.

In the previous sections, we have examined two different sources of error. In Section~\ref{sec:noise}, we examined the impact of charge noise on coherence times and how decoherence peaks at the interdot anticrossing. This effect should be minimized by pulsing across the anticrossing rapidly to minimize the amount of time spent there. In Section~\ref{sec:gates}, we examined the amount of diabatic errors in our system and concluded that we should operate in a regime of large tunnel coupling where we minimize diabatic errors. We find that given the large tunnel coupling, accurate simulations in the transfer process are computationally costly, requiring very fine time resolution. Simulations of unitary errors with a time-dependent detuning as shown in Fig.~\ref{fig:gates}(d) was only possible with the reduced basis, i.e. without excited states, ignoring diabaticity. We note that $T_2^*$ effects discussed in Section~\ref{sec:noise} and diabatic effects discussed in Section~\ref{sec:gates} become appreciable in distinct ranges of ramp rate. Assuming a fixed tunnel coupling of about $400~\text{\textmu{}eV}$ as we have done in the paper, we would have to work in regimes where the ramp rate is approximately $10^5~\text{\textmu{}eV/ns}$  or more before diabatic effects became significant [Fig.~\ref{fig:gates}(b)]. At this point, considering most detuning ranges in a transfer between two quantum dots, the transfer would be concluded in less than 0.1 ns, and therefore $T_2^*$ effects could be considered to be negligible. Conversely, assuming a slower ramp rate, $T_2^*$ effects become more significant but diabatic effects become negligible exponentially.

We would also like to point out that there have been a related analysis performed by Krzywda et al. \cite{krzywda2020adiabatic}. There, the diabaticity effect due to $1/f$ noise in charge transport between dots was analyzed analytically, which is complementary to our investigations into spin transfer process.

Other than tunnel coupling, parameters like spin-orbit coupling strengths and magnetic fields also have an impact on the transport process, but they are either kept constant or are difficult to control in scalable architectures. Recent work also suggests that spin-orbit parameters can be heavily dependent on surface roughness and other characteristics of the device determined during the fabrication process \cite{ruskov2018electron,tanttu2019controlling,ferdous2018interface}.

We have shown here that with a large tunnel coupling and operating at or close to the sweet spot in detuning, it would be possible to overcome these other effects, which is also substantiated by the experimental results obtained \cite{yoneda2021coherent}. Discussing the results of coherent qubit transport in the context of tunnel coupling is also important because it is a highly tunable parameter \cite{eenink2019tunable,zajac2015reconfigurable,takakura2014single} and that will be important in scale-up architectures as well. This is especially the case when we intend to coherently transport qubits in a large-scale structure in the bucket-brigade manner, a scheme that we are primarily concerned with here. In this scheme, we are moving the electrons across multiple dots and therefore it is important to understand the tunneling process, which is where most of the errors occur in our system.

\section{Summary}
\label{sec:conclusion}

In this paper, we adopted an eight-level model for our double quantum dot system, which includes the spin, valley and charge degrees of freedom. We examined two different types of errors in spin transfer, temporal and unitary errors. To understand the temporal errors, we analyzed the spectrum of both $T_2^*$ and $T_2^\mathrm{H}$ times, and estimated that $1/f$ noise with amplitude of $100~\text{\textmu{}}\mathrm{eV}^2/\mathrm{Hz}$ at $1~\mathrm{Hz}$ is able to adequately describe the spectrum of coherence times near the inter-dot anticrossing. As for unitary errors, we model the transport process as $x$- and $z$-rotations on the Bloch sphere by considering only the two lowest-energy states and show that any errors in the form of $x$-rotations are negligible in the rotating frame with large tunnel couplings ($t_\mathrm{c}\gg E_\mathrm{Z}$). Finally, we discussed how the results we presented changes with the size of the tunnel coupling, further cementing its importance.

To conclude, a large tunnel coupling will be key in minimizing the errors that occur during the transport process, especially for spin-flip errors. Also, fast pulsing will be very helpful for avoiding the region of fast dephasing near the inter-dot anticrossing, especially since most errors accumulate near the inter-dot anticrossing. Having coherent qubit transfer in large scale systems will allow for non-local operations, while increasing the inter-connectivity between the qubits.

\begin{acknowledgments}
    We thank I. Hansen, P. Mai, J. Y. Huang, and C. C. Escott for helpful discussions. We acknowledge support from the Australian Research Council (FL190100167 and CE170100012) and the US Army Research Office (W911NF-17-1-0198). The views and conclusions contained in this document are those of the authors and should not be interpreted as representing the official policies, either expressed or implied, of the Army Research Office or the US Government. M.K.F., S.Y., J.D.C. and W.G. acknowledges support from Sydney Quantum Academy. J.Y acknowledges support from JST PRESTO grant (JPMJPR21BA).
\end{acknowledgments}

\appendix

\section{Hamiltonians}
\label{app:hamiltonian}

In Section~\ref{sec:model}, we model our double quantum dot system as an eight-level system, as described by the Hamiltonian given in Eq.~(\ref{eqn:ham8}). Here we explicitly give the matrix representation of the Hamiltonian,
\begin{widetext}
\begin{align}
    \hat{H}_\mathrm{8\times8} = \begin{pmatrix}
    \hat{H}_\mathrm{8\times8}^{(A)} & \hat{H}_\mathrm{8\times8}^{(c)} \\
    \hat{H}_\mathrm{8\times8}^{(c)\dagger} & \hat{H}_\mathrm{8\times8}^{(B)}
    \end{pmatrix}
\end{align}
where each of these terms is a 4-by-4 matrix,
\begin{align}
    \hat{H}_\mathrm{8\times8}^{(A)} = \frac{1}{2}\begin{pmatrix}
    E_\mathrm{Z,A} + (\eta_\mathrm{A} + 1)\varepsilon & 0 & E_\mathrm{v,A} & \Delta_2^{\mathrm{sv}*} \\
    0 & -E_\mathrm{Z,A} - (\eta_\mathrm{A} - 1)\varepsilon & \Delta_1^\mathrm{sv} & E_\mathrm{v,A} \\
    E_\mathrm{v,A}^* & \Delta_1^{\mathrm{sv}*} & E_\mathrm{Z,A} + (\eta_\mathrm{A} + 1)\varepsilon & 0 \\
    \Delta_2^\mathrm{sv} & E_\mathrm{v,A}^* & 0 & -E_\mathrm{Z,A} -(\eta_\mathrm{A} - 1)\varepsilon
    \end{pmatrix}
\end{align}
\begin{align}
    \hat{H}_\mathrm{8\times8}^{(c)} = \hat{H}_\mathrm{8\times8}^{(c)*} = \frac{1}{2}\begin{pmatrix}
    t_\mathrm{c} + t_\mathrm{sd} & t_\mathrm{sf} & 0 & 0 \\
    t_\mathrm{sf} & t_\mathrm{c} - t_\mathrm{sd} & 0 & 0 \\
    0 & 0 & t_\mathrm{c} + t_\mathrm{sd} & t_\mathrm{sf} \\
    0 & 0 & t_\mathrm{sf} & t_\mathrm{c} - t_\mathrm{sd}
    \end{pmatrix}
\end{align}
\begin{align}
    \hat{H}_\mathrm{8\times8}^{(B)} = \frac{1}{2}\begin{pmatrix}
    E_\mathrm{Z,B} + (\eta_\mathrm{B} -1)\varepsilon & 0 & E_\mathrm{v,B} & \Delta_2^{\mathrm{sv}*} \\
    0 & -E_\mathrm{Z,B} - (\eta_\mathrm{B} + 1)\varepsilon & \Delta_1^\mathrm{sv} & E_\mathrm{v,B} \\
    E_\mathrm{v,B}^* & \Delta_1^{\mathrm{sv}*} & E_\mathrm{Z,B} + (\eta_\mathrm{B} - 1)\varepsilon & 0 \\
    \Delta_2^\mathrm{sv} & E_\mathrm{v,B}^* & 0 & -E_\mathrm{Z,B} - (\eta_\mathrm{B} + 1)\varepsilon
    \end{pmatrix}
\end{align}
in the basis of,
\begin{align}
    \{\ket{A,\uparrow,-k_0},\ket{A,\downarrow,-k_0},\ket{A,\uparrow,+k_0},\ket{A,\downarrow,+k_0},\ket{B,\uparrow,-k_0},\ket{B,\downarrow,-k_0},\ket{B,\uparrow,+k_0},\ket{B,\downarrow,+k_0}\}
\end{align}
\end{widetext}
Here, $E_{\mathrm{Z},i}$ refers to the Zeeman splitting, $\eta_i$ is the Stark shift, $\varepsilon$ is the detuning of the double dot, $t_\mathrm{c}$ is the tunnel coupling between the dots, $t_\mathrm{sd}$ and $t_\mathrm{sf}$ are respectively the spin-dependent and spin-flip tunneling parameters arising from spin-orbit coupling, $\Delta_{1(2)}^{\mathrm{sv}}$ are the spin-valley coupling terms, and finally $E_{\mathrm{v},i}$ are the valley splitting terms. 

We note here that we consider the coupling terms to be completely real, taking only the magnitude of $t_\mathrm{c}$, $t_\mathrm{sd}$, and $t_\mathrm{sf}$. This is because the imaginary part of these terms only causes the electron to gain a phase as it moves across the dots, which is difficult to differentiate experimentally from other contributions (\textit{i.e.} inter-dot qubit frequency difference).

In Section~\ref{sec:noise}, we discussed the role of $1/f$ noise in the system in the context of the four-level model. Here, we explicitly show the matrix representation of that model Hamiltonian in the basis of $\left\{\ket{A,\uparrow},\ket{A,\downarrow},\ket{B,\uparrow},\ket{B,\downarrow}\right\}$ where we consider only the lower valley eigenstates, and the four-level Hamiltonian can be obtained by considering only the intra-valley components of the full $8\times8$ Hamiltonian.
\begin{widetext}
\begin{align}
    \hat{H}_{4\times 4} = \frac{1}{2}\begin{pmatrix}
    E_\mathrm{Z,A} + \left(\eta_\mathrm{A} + 1\right)\varepsilon & 0 & t_\mathrm{c} + t_\mathrm{sd} & t_\mathrm{sf} \\
    0 & -E_\mathrm{Z,A} - \left(\eta_\mathrm{A} - 1\right)\varepsilon & t_\mathrm{sf} & t_\mathrm{c} - t_\mathrm{sd} \\
    t_\mathrm{c} + t_\mathrm{sd} & t_\mathrm{sf} & E_\mathrm{Z,B} + \left(\eta_\mathrm{B} - 1\right)\varepsilon & 0 \\
    t_\mathrm{sf} & t_\mathrm{c} - t_\mathrm{sd} & 0 & -E_\mathrm{Z,B} - \left(\eta_\mathrm{B} + 1\right)\varepsilon
    \end{pmatrix}\:.
\end{align}
\end{widetext}

In Section~\ref{sec:gates}, we further simplified the model and performed a Schrieffer-Wolff transformation \cite{winkler2003spin}. We consider the four-level model Hamiltonian to be a sum of $\hat{H}_0$ and $V$ where $\hat{H}_0$ is given as,
\begin{align}
    \hat{H}_0 = \frac{1}{2}\begin{pmatrix}
    E_\mathrm{Z} + \varepsilon & 0 & t_\mathrm{c} & 0 \\
    0 & -E_\mathrm{Z} + \varepsilon & 0 & t_\mathrm{c} \\
    t_\mathrm{c} & 0 & E_\mathrm{Z} - \varepsilon & 0 \\
    0 & t_\mathrm{c} & 0 & -E_\mathrm{Z} -\varepsilon
    \end{pmatrix}~,
\end{align}
and the perturbation term $V$ is given as,
\begin{align}
    V = \frac{1}{2}\begin{pmatrix}
    \eta_A\varepsilon & 0 & t_\mathrm{sd} & t_\mathrm{sf} \\
    0 & -\eta_A\varepsilon & t_\mathrm{sf} & -t_\mathrm{sd} \\
    t_\mathrm{sd} & t_\mathrm{sf} & \eta_B\varepsilon -\Delta E_\mathrm{Z} & 0 \\
    t_\mathrm{sf} & -t_\mathrm{sd} & 0 & -\eta_B\varepsilon +\Delta E_\mathrm{Z}
    \end{pmatrix} \:.
\end{align}
where $\hat{H}_0$ is non-diagonal since $t_\mathrm{c}$ is large with respect to the Zeeman splitting $E_\mathrm{Z}$, and therefore cannot be treated as perturbation. Performing the Schrieffer-Wolff transformation, we obtain an effective $2\times2$ Hamiltonian,
\begin{align}
    H_\mathrm{2\times 2} = \frac{1}{2}\hbar\omega_0\mathbb{1} +  \frac{1}{2}\hbar\omega\sigma_z + \frac{1}{2}\hbar g \sigma_x
\end{align}
where
\begin{align}
    \hbar\omega_0 \approx -\sqrt{t_\mathrm{c}^2+\varepsilon^2}\;,
\end{align}
\begin{multline}
    \hbar\omega \approx -\frac{E_\mathrm{Z,A} + E_\mathrm{Z,B} + \varepsilon(\eta_\mathrm{A} + \eta_\mathrm{B})}{2} + \\ 
    \hspace{4.4mm}\frac{2t_\mathrm{c}t_\mathrm{sd}+\Delta E_\mathrm{Z}\varepsilon + \varepsilon^2(\eta_\mathrm{A}-\eta_\mathrm{B})}{2\sqrt{t_\mathrm{c}^2+\varepsilon^2}}\;,
\end{multline}
and 
\begin{align}
    \hbar g \approx -\frac{t_\mathrm{c} t_\mathrm{sf}}{\sqrt{t_\mathrm{c}^2+\varepsilon^2}}~.
\end{align}
We have neglected second order terms and higher. The main feature of this expression lies in the $\sigma_x$ term which is proportional to $t_\mathrm{sf}$, indicating that any errors in a form of an out-of-plane rotation depend on the magnitude of $t_\mathrm{sf}$. Furthermore, in the rotating frame, the off-diagonal term proportional to $\sigma_x$ averages to zero due to a rapidly rotating term at the frequency of the rotating frame (which is the precession frequency at $\varepsilon=-1.35$). The Hamiltonian parameters used for the calculations in the rest of this document are derived from a combination of experimental parameters \cite{yoneda2021coherent} as well as what is known in the literature \cite{huang2014spin,wang2013charge,zhang2020giant,yang2013spin}. Most of the valley related parameters are treated as free parameters that we can tune to fit the spin probabilities as obtained from the experiment.
\vspace{1mm}
\begin{widetext}
\begin{center}
\begin{longtable}{| c | c | c |}
    \hline
    Zeeman splitting (1T) & $E_\mathrm{Z}$ & 116 \textmu{}eV (28 GHz) \\\hline
    Zeeman splitting difference between dots (1T) & $\Delta E_\mathrm{Z}$ & 138 $\mathrm{neV}$ (33.4 MHz)  \\\hline
    Zeeman splitting (1.42T) & $E_\mathrm{Z}$ & 166 \textmu{}eV (40 GHz) \\\hline
    \hspace{1mm} Zeeman splitting difference between dots (1.42T) \hspace{1mm} & $\Delta E_\mathrm{Z}$ & 205 $\mathrm{neV}$ (49.4 MHz) \\\hline
    Valley splitting in dot A & $E_\mathrm{v,A}$ & 330 \textmu{}eV (80 GHz) \\\hline
    Valley splitting in dot B & $E_\mathrm{v,B}$ & 330 \textmu{}eV (80 GHz) \\\hline
    Valley phase in dot A & $\phi_\mathrm{A}$ & 0 rad \\\hline
    Valley phase in dot B & $\phi_\mathrm{B}$ & 0 rad \\\hline
    Spin valley mixing, $\expval{\downarrow,v^\prime|H_{so}|\uparrow,v}$ & $\Delta_1^\mathrm{sv}$ & 20.7 $\mathrm{neV}$ (5 MHz) \\\hline
    Spin valley mixing, $\expval{\uparrow,v^\prime|H_{so}|\downarrow,v}$ & $\Delta_2^\mathrm{sv}$ & 166 $\mathrm{neV}$ (40 MHz) \\\hline
    Tunnel coupling & $t_\mathrm{c}$ & 430 \textmu{}eV (104 GHz) \\\hline
    Spin-dependent tunnel coupling & $t_\mathrm{sd}$ & -14.1 $\mathrm{neV}$ (-3.4 MHz) \\\hline
    Spin-flip tunnel coupling & $t_\mathrm{sf}$ & -14.1 $\mathrm{neV}$ (-3.4 MHz) \\\hline
    Stark shift in dot A & $\eta_\mathrm{A}$ & 186 MHz/eV (0.77 MHz/THz) \\\hline
    Stark shift in dot B & $\eta_\mathrm{B}$ & -33.8 MHz/eV (-0.14 MHz/THz) \\
    \hline
    \caption{Hamiltonian parameters used in numerical simulations.}
\end{longtable}
\end{center}
\end{widetext}

We verify the accuracy of these parameters as well as the consistency of the four-level model and the two-level model after Schrieffer-Wolff transformation by checking the qubit frequency of the double dot system using these reduced models.

\begin{figure}[!ht]
    \centering
    \includegraphics[width=0.45\textwidth]{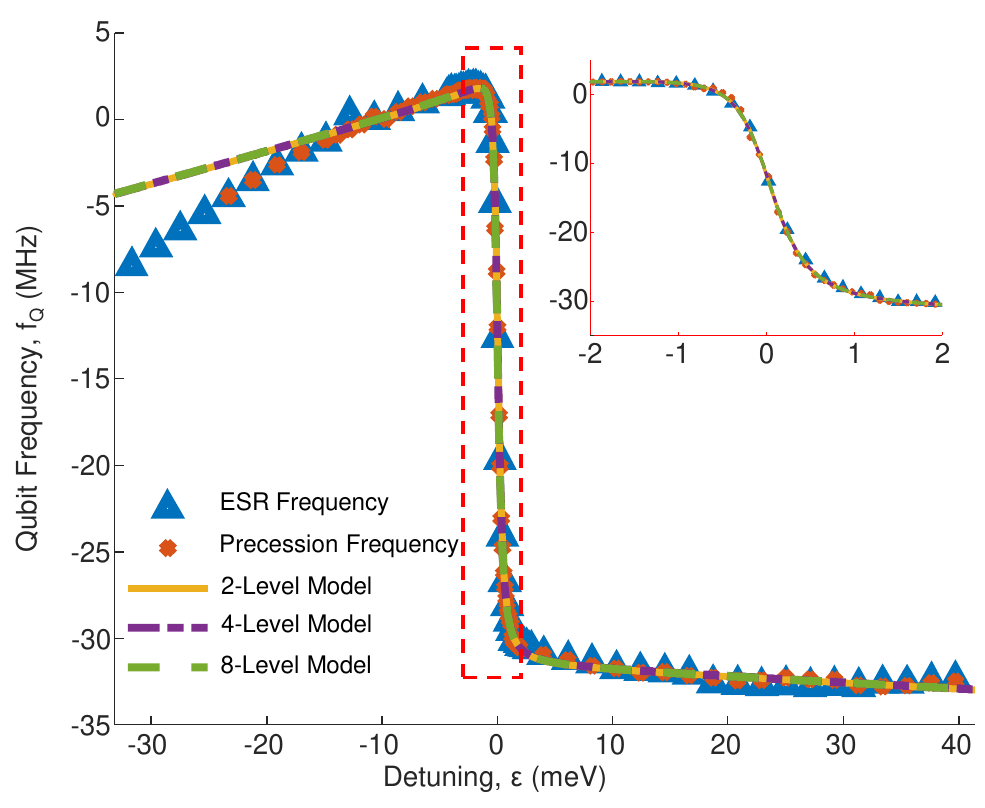}
    \caption{Verification of the different models' accuracy in reproducing the experimental qubit frequency, $f_\mathrm{Q}$. The inset shows a zoom-in of the fit within a smaller detuning range.}
    \label{appfig:qubitfreq}
\end{figure}

We show that for the given parameters, all the three models match well the experimental results (ESR and precession frequency), across the relevant range of detuning values. We show in the inset the fitting near the anticrossing to show that it is also well-fitted for small detuning values.

\section{Simulation of Noise}
\label{app:1fsimulation}

In Section~\ref{sec:noise}, we examined the spectrum of $T_2^*$ and $T_2^\mathrm{H}$ as a function of detuning by the addition of computer generated noise to the qubit Hamiltonian, as it undergoes time evolution. We have also outlined how we obtained the spectrum of decoherence rates. In this section, we detail how we generated the $1/f$ noise spectrum.

We begin by generating an array of $10^5$ random numbers, $x(t)$. At this stage, the array is agnostic to the time range, and the range of $t_\mathrm{evol}$ that it crosses is dependent on the time steps $\Delta t$ used in the simulation. We perform discrete Fourier transformation on this array to convert it from the time domain into the frequency domain,
\begin{align*}
    x(f) = \sum_{t=1}^{N-1} x(t) e^{-2i\pi t f/N} \:.
\end{align*}
The range of frequencies that we will consider include negative frequencies, and we can scale the $x(f)$ by a scaling factor given as,
\begin{align*}
    A(f) = |\delta A|f^{-1/2}
\end{align*}
where the frequency is scaled by a factor of $-1/2$ because we consider $1/f$ noise and the square root of the power spectrum is inversely proportional to $\sqrt{f}$. The amplitude in the scaling factor $|\delta A|$ is a coefficient that is determined such that the resulting PSD extends to $100~\text{\textmu{}}\mathrm{eV}^2/\mathrm{Hz}$ at $1~\mathrm{Hz}$, and will change with the time step $\Delta t$ such that the ratio of $|\Delta A|^2$ to $\Delta t$ remains the same. We then perform an inverse Fourier transform and obtain the scaled time domain spectrum which can be used as noise input for the simulations.

Examples of power spectral densities of these noise spectra are shown in Fig.~\ref{fig:error}(c) in blue and red. In general, we simulate noise spectra with time steps that vary from $0.05~\mathrm{ns}$ to $0.05~\mathrm{s}$ and have lower cutoff frequencies given by $f_l=1/(N\Delta t)$ which depends on both the time step and the number of samples. The higher cutoff frequency, on the other hand, is given by, $f_h=1/(2\Delta t)$ and depends only on the size of the time steps.

\section{Numerical Simulation of Coherence Times}
\label{app:numerics}



\begin{figure}[!ht]
    \centering
    \includegraphics[width=0.45\textwidth]{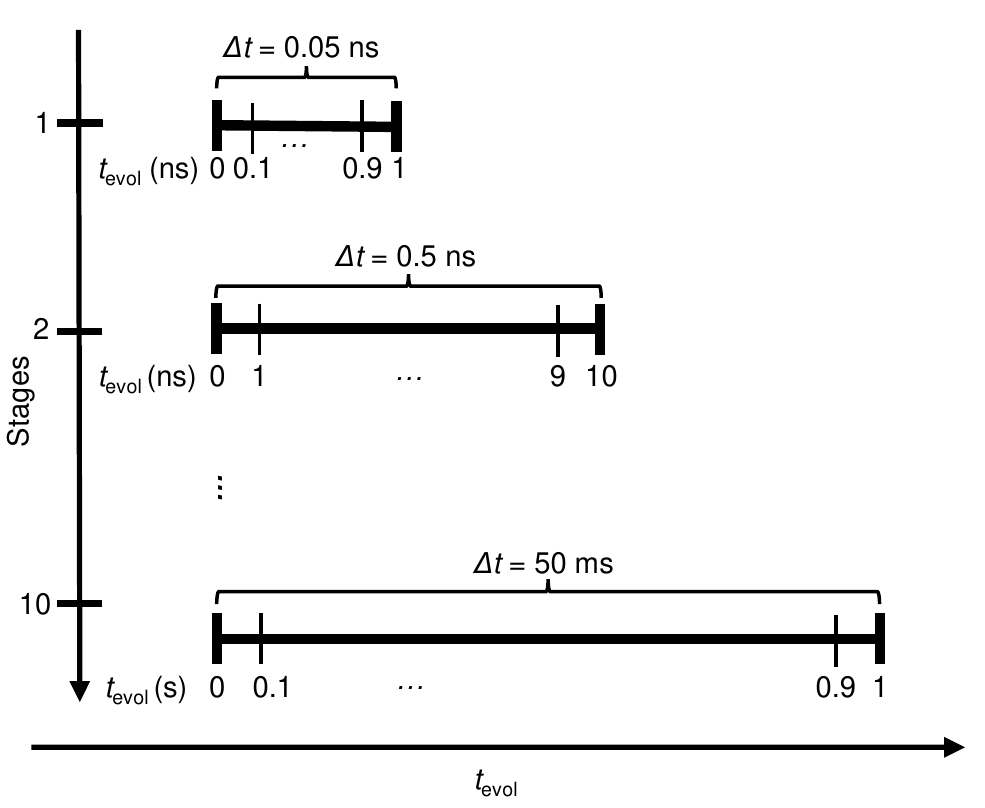}
    \caption{Schematic of numerical calculation of coherence times. The calculations are partitioned into different sections with varying time steps $\Delta t$ and evolution times $t_\mathrm{evol}$.}
    \label{appfig:numerics}
\end{figure}

In our simulations, we need to probe coherence times ranging across several orders of magnitude. In order to do so effectively, we partition the simulation into different sections, each with different time steps, $\Delta t$, and total evolution times, $t_\mathrm{evol}$. We show the schematic of how that is performed in Fig.~\ref{appfig:numerics}. We partition each section such that the parameters $\Delta t$ and $t_\mathrm{evol}$ differ no more than 2 orders of magnitude, thus ensuring that the simulation time remains manageable.

As shown in Fig.~\ref{appfig:numerics}, this allows us to obtain results for the smallest $t_\mathrm{evol}$ of 0.1 ns up to 1 s. Each section of the simulation also has a different noise spectrum since $\Delta t$ is changing from section to section. To maintain consistency between the runs, the amplitude pre-factor $|\delta A|$ for the noise also changes such that the resulting PSD amplitude remains the same. Examples of these generated noise spectra with different $\Delta t$ are shown in Fig.~\ref{fig:error}(c).

\section{SET Spectroscopy}
\label{app:setspec}

The noise spectrum shown in Fig.~\ref{fig:error}(b) was measured with the single electron transistor (SET) in the device used for the spin transport experiments \cite{yoneda2021coherent}. Obtaining the SET noise spectrum is a two step process. In the first step, we measured time traces of the fluctuations in the current, $I(t)$, at different top gate voltage values, $V_i$, as we sweep across a SET peak. Secondly, in order to obtain the corresponding trace for fluctuations in voltage, we will need to also measure the sensitivity at these voltage values.

We choose two sensitivity points at $\pm \delta V$ away from the chosen $V_i$ where $I(t)$ is measured, with $\pm \delta V$ small enough such that SET will give an approximately linear response within that range. At each sensitivity point, fluctuations in the current, $I(t)$, is sampled with 10 repetitions of 20 second-long current signals. We define the ratio of $\delta\bar{I}$ to $2\delta V$ as the sensitivity of the SET, where $\delta\bar{I}$ is the difference of the mean current between $V_i \pm \delta V$. The sensitivity can be used to convert the current response, $I(t)$, to voltage response, $V(t)$ for each top gate voltage, $V_i$. The fast Fourier transform of the resulting $V(t)$ scaled by a leverarm factor ($\sim0.2~\mathrm{eV/V}$) corresponds to the PSD shown in Fig.~\ref{fig:error}(c) (purple trace).

This measurement will be able to capture the fluctuations from the environment around the SET, and we showed in our studies that these fluctuations in the current through the SET is also a good indicator of the charge fluctuations on the qubit gates in the form of detuning noise. This particular measurement of noise only measures up to frequencies of $10~\mathrm{kHz}$ due to the filters in the setup, so it does not directly inform us of the noise levels at high frequencies (on the order of megahertz) which is the order of time on which the qubit is being operated. However, it would be sensible to use the amplitude of noise as given by this measurement as an indication of the level of noise expected at all frequencies, and so we extrapolate the noise spectrum to higher frequencies, obtaining the spectra shown in Fig.~\ref{fig:error}(c). The missing piece of the puzzle in determining the expected noise spectrum is the slope of the noise spectrum. It is generally expected that the source of charge noise is two-level fluctuators each with noise statistic $1/f^2$ and they would average out to a $1/f$ spectrum \cite{paladino20141,chan2018assessment}. Therefore, we take the noise level at $1~\mathrm{Hz}$ as the point of reference and assume the actual noise spectrum extends with a slope of $1/f$. The slope, as shown in the experimental noise spectroscopy, can be much steeper than $1/f$ especially at low frequencies, but we do not interpret it to be indicative of the noise characteristic across all frequencies. This is because in a nanostructure, noise can be dominated by only a few nearby two-level fluctuators at certain frequency ranges, and it is common that they do not average out to an exact $1/f$ noise when observed across a narrow frequency range.

The SET spectroscopy data is valuable in understanding the amplitude of the electrical noise present in the system, by probing directly the charge fluctuations via the SET, and will be useful especially for probing low frequency charge noise \cite{connors2019low}. It is also a fast measurement that can be used as a benchmarking tool for noise in qubit systems, without performing a full qubit noise spectroscopy \cite{connors2022charge}. Although it is not capable of probing noise in the high frequency regime, the results from the studies here indicate that extrapolating the low frequency results to higher frequencies is consistent with the qubit coherence results, and therefore remains a useful predictive tool.

\section{Sweet Spots in Detuning}
\label{app:sweetspot}

In Sec. III, we showed that a coherence sweet spot in detuning can occur as a result of the qubit dispersion going to zero in dot A, due to the Stark shift from the detuning between dots causing at first a decrease in spin frequency and then a sudden increase when the spin finally tunnels through. Whether a particular interdot transition will have these properties is a random feature that depends on the particular roughness profile of the Si/SiO$_\text{2}$ interface between those two dots. Therefore, for a single pair of quantum dots, there is a 75\% chance of finding a sweet spot on either side of the anticrossing (ignoring correlation between Stark shift slope and interdot transition frequency shift). Therefore, when it comes to a chain of dots, it becomes increasingly unlikely that no sweet spots will be available across a number of transitions.

Upon closer inspection of the results, the sweet spot in both simulation and experiment can be seen from the results of experiment 1 in Fig.~\ref{fig:error}(e) (blue trace), but that is not the case for experiment 2 which we plotted in red.

Both of these experiments (1 and 2) were Ramsey-like experiments where the amplitude of Ramsey fringes is measured as a function of detuning and evolution time. By extracting the decay of the fringes, the coherence times can be obtained. But there exists a slight difference in the protocol for these two experiments, and that is the order in which the sweep over evolution time and the averaging over the fringe phase shifts are performed. In experiment 1, the fringe amplitudes are measured in a sweep over time before repeating the results for different phase shifts and averaging over the phases. However, in experiment 2, the reverse process is executed, where the amplitude is measured for different phase shifts before varying the evolution time. However, we believe that these changes cannot explain the difference between the two sets of measured $T_2^\mathrm{H}$.

The difference between these data sets can occur due to a small ($\sim0.5~\mathrm{meV}$) shift in the detuning of experiment 1 [shown by the blue trace in Fig.~\ref{fig:error}(d)]. The absence of the sweet spot may also be due to a detuning jump between detuning steps during experiment 2 [shown by the red trace in Fig.~\ref{fig:error}(d)]. These detuning jumps can occur due to charge noise in the system, and is difficult to track during the course of the experiment. It is also plausible that the sweet spot might be less distinct in the experimental data sets due to another source of noise that changes the noise spectral density around 10 kHz (i.e. $>1/10^2$ \textmu{}s), which limits $T_2^\mathrm{Hahn}$ times (despite the fact that $T_2^*$ times seem fully accounted for).




\section{Effect of Magnetic Fields on Decoherence Rates}
\label{app:noise}

\begin{figure}[!ht]
    \centering
    \includegraphics[width=0.45\textwidth]{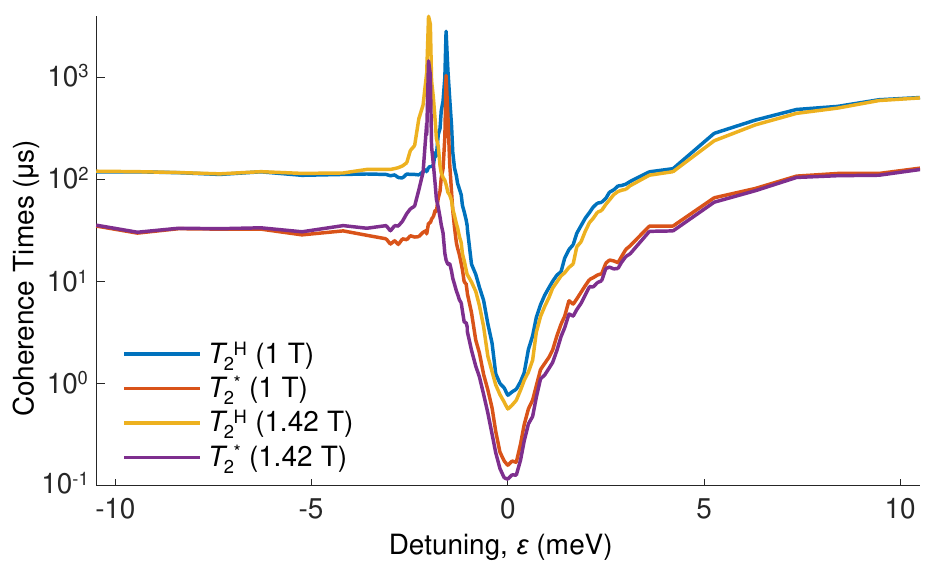}
    \caption{Coherence times at different magnetic fields. We show the different coherence times, $T_2^*$ and $T_2^\mathrm{H}$, at different magnetic fields of $1~\mathrm{T}$ and $1.42~\mathrm{T}$.}
    \label{appfig:coherence}
\end{figure}

In Section~\ref{sec:noise}, we discussed the coherence times in the system and showed that $1/f$ noise was able to sufficiently account for the detuning dependence of coherence times observed in the experiment conducted in Ref.~(\onlinecite{yoneda2021coherent}). In addition, we also showed that an echo sequence improves the coherence times, giving us a long $T_2^\mathrm{H}$, but at a different magnetic field. Here, we plot both the Ramsey coherence time $T_2^*$ and the Hahn echo time $T_2^\mathrm{H}$ at both magnetic fields of $1~\mathrm{T}$ and $1.42~\mathrm{T}$, showing that the effect of the echo sequence outweighs the effect of the drop in magnetic field, thus supporting the result shown in Fig.~\ref{fig:error}, where approximately an order of magnitude improvement in coherence times was observed with the Hahn echo.

\section{Adiabaticity}
\label{app:adiabaticity}

\begin{figure}[ht]
    \centering
    \includegraphics[width=0.45\textwidth]{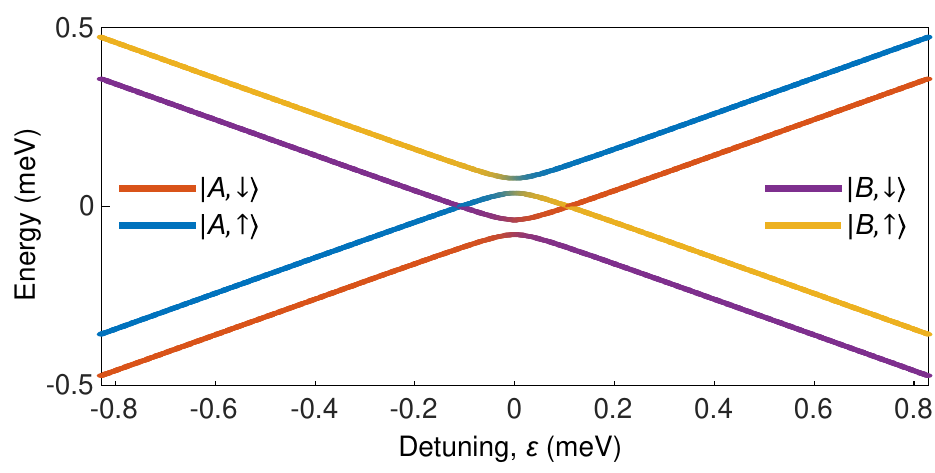}
    \caption{Four-level energy diagram with $t_\mathrm{c} < E_\mathrm{z}$. We show here the energy level diagram for $t_\mathrm{c} = 41~\text{\textmu{}}\mathrm{eV} \approx h\times10~\mathrm{GHz}$ at 1 T with $E_\mathrm{z} = 116~\text{\textmu{}}\mathrm{eV} \approx h\times28~\mathrm{GHz}$.}
    \label{appfig:diabatic}
\end{figure}

In the results of this paper, we have considered diabatic effects to be negligible in the transfer process, since we take the tunnel coupling to be much larger than the Zeeman splitting, $t_\mathrm{c} \gg E_\mathrm{z}$. However, there are in fact two regimes that concern us in spin transport, one with $t_\mathrm{c} \gg E_\mathrm{z}$ represented in Fig.~\ref{fig:error}(a), which is the main regime of concern in this paper and another with $t_\mathrm{c} < E_\mathrm{z}$ shown here in Fig.~\ref{appfig:diabatic}, where there are several state leakage pathways. In the $t_\mathrm{c} < E_\mathrm{z}$ regime, state leakage becomes significant because of the additional anticrossings to the left and right of the inter-dot anticrossing.

However, it is also worth noting that having a small tunnel coupling does not mean that coherent spin transport cannot be performed. There remain several ways to minimize diabatic errors through manipulation of the pulses. We can either calculate a single ramp rate or have a pulse with variable ramp rates that change with the detuning, $\varepsilon$. Either of these methods should allow us to cross the spin-flip anticrossing diabatically but the inter-dot anticrossing adiabatically, ensuring that the qubit remains in the desired state during the transfer \cite{buonacorsi2020simulated,krzywda2020adiabatic}. At the same time we avoid the region of large Stark shift between the two spin-flip anticrossings. However, the ramp rate in either of these methods will be limited by the size of the inter-dot anticrossing since we are in the regime of small tunnel coupling, which may lead to increased time spent at the inter-dot anticrossing, thus increasing errors.

\section{Transfer Process as an x Rotation}
\label{app:xrot}

\begin{figure}[!ht]
    \centering
    \includegraphics[width=0.45\textwidth]{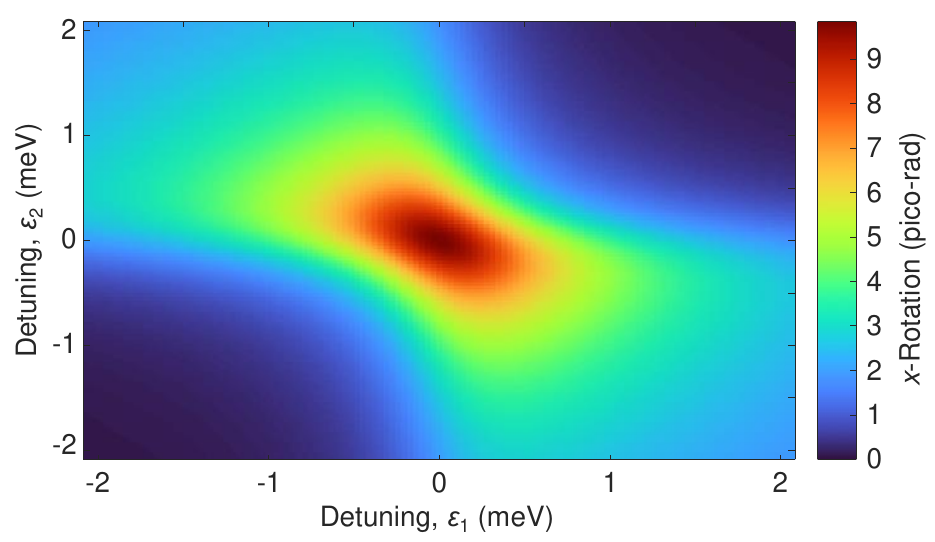}
    \caption{Effective $x$ gate after transfer. Figure shows the magnitude of effective $x$-rotations after a transfer from $\varepsilon_1$ to $\varepsilon_2$ over 8 ns.}
    \label{appfig:xrot}
\end{figure}

In Sec.~\ref{sec:gates}, we examined the transfer process as effective rotations by considering the Hamiltonian in the effective basis of the ground spin up and down basis. We show here the results of characterizing the $x$-rotation by looking at the coefficient of the $\sigma_x$ component of the effective rotation obtained from the logarithm of the effective unitary operator. In Fig.~\ref{appfig:xrot}, we show how the $\sigma_x$ component changes with start ($\varepsilon_1$) and end ($\varepsilon_2$) detuning positions. 

Given that Fig.~\ref{appfig:xrot} is plotted for fixed ramp time, we notice that near the center of the plot where both the start and end detuning positions ($\varepsilon_1$ and $\varepsilon_2$ respectively) are close to zero, the amount of $x$-rotation is greater, indicating for a small ramp time, the amount of $x$-rotation is influenced by the amount of time spent close to the anticrossing, such that the qubit gains more errors by spending more time near the inter-dot anticrossing. When the dot movement occurs only within the same dot (top right and bottom left quadrants of the figure), the error remains small because the electron does not cross the inter-dot anticrossing. When the transfer is across a larger range of detuning, corresponding to the top left and bottom right regions of the figure, we observe that there is now a larger error since the electron is being pulsed across the double dot, therefore spending time at the inter-dot anticrossing. The error is largest in magnitude near the inter-dot anticrossing, regardless of the start and end detuning points, because the electron is always kept near the anticrossing and accumulates spin-flip errors.

\section{Role of Spin-Orbit Coupling}
\label{app:spinorbit}

In SiMOS devices, spin-orbit coupling plays a very important role and has contributions from both what is traditionally known as Rashba and Dresselhaus effects \cite{jock2018silicon,veldhorst2015spin}. Rashba effects arise from structural inversion asymmetry due to the axial strain in the [001] crystallographic direction. On the other hand, Dresselhaus effects exist because the quantum dots are situated close to the silicon-silicon dioxide interface, and therefore has bulk inversion symmetry as well. We can model the spin-orbit Hamiltonian as \cite{zhao2018coherent,huang2017electrically},
\begin{align}
    H_\mathrm{soc} = \alpha_\mathrm{R} \left(k_x\sigma_y - k_y\sigma_x\right) + \beta_\mathrm{D} \left(k_y\sigma_y - k_x\sigma_x\right)\:,
\end{align}
where $\alpha_\mathrm{R}$ and $\beta_\mathrm{D}$ are the Rashba and Dresselhaus coefficients respectively. We define the $\sigma_x$ and $\sigma_y$ to be directions along the crystallographic axes of the silicon lattice.

In the most current model of spin-orbit coupling \cite{zhao2018coherent,huang2017electrically}, these effects combine to give rise to spin-orbit induced tunnel couplings, which are accounted for in our system in the form of spin-dependent and spin-flip tunnel couplings. Spin-dependent tunnel couplings $t_\mathrm{sd}$ changes the inter-dot tunnel coupling such that we have different tunnel couplings depending on whether the spin is in the up or down state. Spin-flip tunnel coupling $t_\mathrm{sf}$ on the other hand couples opposite spin states across different dots, \textit{i.e.} between $\ket{A,\uparrow}$ and $\ket{B,\downarrow}$. As a result, the spin-orbit Hamiltonian in the qubit quantization axis is of the following form,
\begin{align}
    H_\mathrm{soc} = \frac{1}{2}\begin{pmatrix} t_\mathrm{sd} & t_\mathrm{sf} \\ t_\mathrm{sf} & -t_\mathrm{sd} \end{pmatrix}
\end{align}
We show here that we have both diagonal and off-diagonal spin-orbit terms in the qubit Hamiltonian, with the diagonal terms of $t_\mathrm{sd}$ being atypical in spin-orbit coupling, but occurs due to the spin-orbit Hamiltonian being defined along the crystallographic axis but the qubit Hamiltonian being defined by the in-plane external magnetic field, which results in these two Hamiltonians being orthogonal to each other \cite{huang2017electrically}. By aligning the spin-orbit Hamiltonian into the quantization axis of the qubit Hamiltonian, the spin-orbit terms are rotated, resulting in both diagonal and off-diagonal spin-orbit terms, as shown below, if we assume an in-plane rotation of $\pi/4$ and an out-of-plane rotation of $\pi/2$,
\begin{multline}
    H_\mathrm{soc}^\prime = (\alpha_\mathrm{R}-\beta_\mathrm{D})(k_\mathrm{x}-k_\mathrm{y})\sigma_z \\
    - \frac{1}{2}(\alpha_\mathrm{R}+\beta_\mathrm{D})(k_\mathrm{x}+k_\mathrm{y})(\sigma_x - \sigma_y)
\end{multline}

In particular, we were able to determine the spin-dependent tunnel coupling from a fitting of the qubit frequency spectrum. The spin-flip tunnel coupling, while present in the four-level model of the Hamiltonian, does not have an impact on the qubit frequency spectrum in the rotating frame, because off-diagonal terms average out to zero in the rotating approximation. In the simulations, we assume the spin-flip tunnel coupling to be equal to the spin-dependent tunnel coupling. This is because both of the spin-dependent and spin-flip terms originate from both Rashba and Dresselhaus effects and will be a function of $\alpha_\mathrm{R}$ and $\beta_\mathrm{D}$, up to a trigonometric factor which arises from the rotation of the spin-orbit Hamiltonian into the qubit quantization axis. Therefore, the spin-dependent tunnel coupling, $t_\mathrm{sd}$, and spin-flip tunnel coupling, $t_\mathrm{sf}$, are expected to be of the same order of magnitude.



\bibliographystyle{apsrev4-2}
\bibliography{ct_refs}

\end{document}